\documentclass[prb,superscriptaddress,twocolumn,floatfix]{revtex4-1}
\usepackage{graphicx}
\usepackage{amsmath,amssymb}

\def\bscco{Bi$_2$Sr$_2$CaCu$_2$O$_{8+\delta}$}

\def\lsco{La$_{2-x}$Sr$_x$CuO$_4$}
\def\lbco{La$_{2-x}$Ba$_x$CuO$_4$}

\def\ybco{YBa$_2$Cu$_3$O$_{6+x}$}

\def\bscco{Bi$_2$Sr$_2$CaCu$_2$O$_{8+\delta}$}

\newfont{\fc}{cmssbx10 scaled 1000}

\begin{document}

\title{Uniaxial linear resistivity of superconducting La$_{1.905}$Ba$_{0.095}$CuO$_4$ induced by an external magnetic field}

\author{Jinsheng Wen}
\author{Qing Jie}
\affiliation{Condensed Matter Physics \&\ Materials Science Department, Brookhaven National Laboratory, Upton, NY 11973-5000, USA} 
\affiliation{Department of Materials Science and Engineering,
Stony Brook University, Stony Brook, New York 11794, USA}
\author{Qiang Li}
\author{M. H\"ucker}
\affiliation{Condensed Matter Physics \&\ Materials Science Department, Brookhaven National Laboratory, Upton, NY 11973-5000, USA} 
\author{M. v.\ Zimmermann}
\affiliation{Hamburger Synchrotronstrahlungslabor (HASYLAB) at Deutsches Elektronensynchrotron (DESY), 22603 Hamburg, Germany} 
\author{Su Jung Han}
\affiliation{Condensed Matter Physics \&\ Materials Science Department, Brookhaven National Laboratory, Upton, NY 11973-5000, USA} 
\affiliation{Department of Materials Science and Engineering,
Stony Brook University, Stony Brook, New York 11794, USA}
\author{Zhijun Xu}
\affiliation{Condensed Matter Physics \&\ Materials Science Department, Brookhaven National Laboratory, Upton, NY 11973-5000, USA}  
\affiliation{Physics Department, The City College of New York,
New York, NY 10031, USA}
\author{D. K. Singh}
\affiliation{Department of Materials Science and Engineering, University of Maryland, College Park, MD 20742, USA}
\affiliation{NIST Center for Neutron Research, National Institute of
Standards and Technology, Gaithersburg, Maryland 20899, USA}
\author{R. M. Konik}
\author{Liyuan Zhang}
\author{Genda Gu}
\author{J. M. Tranquada}
\affiliation{Condensed Matter Physics \&\ Materials Science Department, Brookhaven National Laboratory, Upton, NY 11973-5000, USA}  
\date{\today}

\begin{abstract}
We present an experimental study of the anisotropic resistivity of superconducting \lbco\ with $x=0.095$ and transition temperature $T_c=32$~K.  In a magnetic field perpendicular to the CuO$_2$ layers, $H_\bot$, we observe that the resistivity perpendicular to the layers, $\rho_\bot$, becomes finite at a temperature consistent with previous studies on very similar materials; however, the onset of finite parallel resistivity, $\rho_\|$, occurs at a much higher temperature.  This behavior contradicts conventional theory, which predicts that $\rho_\bot$ and $\rho_\|$ should become finite at the same temperature.  Voltage vs.\ current measurements near the threshold of voltage detectability indicate linear behavior perpendicular to the layers, becoming nonlinear at higher currents, while the behavior is nonlinear from the onset parallel to the layers.  These results, in the presence of moderate $H_\bot$, appear consistent with superconducting order parallel to the layers with voltage fluctuations between the layers due to thermal noise.   In search of uncommon effects that might help to explain this behavior, we have performed diffraction measurements that provide evidence for $H_\bot$-induced charge and spin stripe order.   The field-induced decoupling of superconducting layers is similar to the decoupled phase observed previously in \lbco\ with $x=1/8$ in zero field.
\end{abstract}

\maketitle

\section{Introduction}

The behavior of underdoped cuprate superconductors in strong magnetic fields, especially fields applied perpendicular to the CuO$_2$ planes, has been of interest in recent years with regard to observations of quantum oscillations.\cite{tail09,seba12}   There are questions as to how the superconducting order is destroyed and how the high-field, low-temperature ``normal'' state compares with the pseudogap phase found in zero field at $T>T_c$.  While the quantum oscillation observations in the underdoped regime are largely limited to the \ybco\ system, the questions regarding the high-field, low-temperature phase are of relevance to all hole-doped cuprates.

The theory for the destruction of superconducting order in layered systems by $H_\bot$ is fairly well established.   Superconducting order within the CuO$_2$ planes is stabilized by Josephson coupling between the layers.\cite{klei92,uema98}  In the mixed phase, the vortex lines behave like stacks of two-dimensional pancake vortices.\cite{clem91}  Weak pinning of vortices by disorder tends to cause the pancake vortices to wander and become misaligned between layers, thus reducing the effective Josephson coupling.\cite{glaz91,daem93,bula95,kosh96b}   When the Josephson coupling is small enough, fluctuations of the phase of the superconducting order parameter may occur, in which case the resistivity becomes finite.   As two-dimensional superconductivity cannot survive at any significant $H_\bot$,\cite{fish91} $\rho_\|$ is expected to become finite at the same time as $\rho_\bot$.\cite{kosh96a,ogan06}  Thus, phase fluctuations, including vortex motion, limit the regime of zero resistivity.

An effective probe of the Josephson coupling and its field dependence is the Josephson plasma resonance, measured by infrared reflectivity.\cite{bula95}  Experimental measurements on optimally and over-doped cuprates yield a field dependence of the effective Josephson coupling in good agreement with theoretical predictions.\cite{duli01,scha10}  In the case of underdoped \lsco\ (LSCO), however, Schafgans {\it et al.}\cite{scha10} found that the coupling dropped more rapidly than predicted, and they proposed that this might be associated with field-induced spin-stripe order.\cite{lake02,chan08,fuji12} 

\begin{figure}[bt]
\centerline{\includegraphics[width=3.3in]{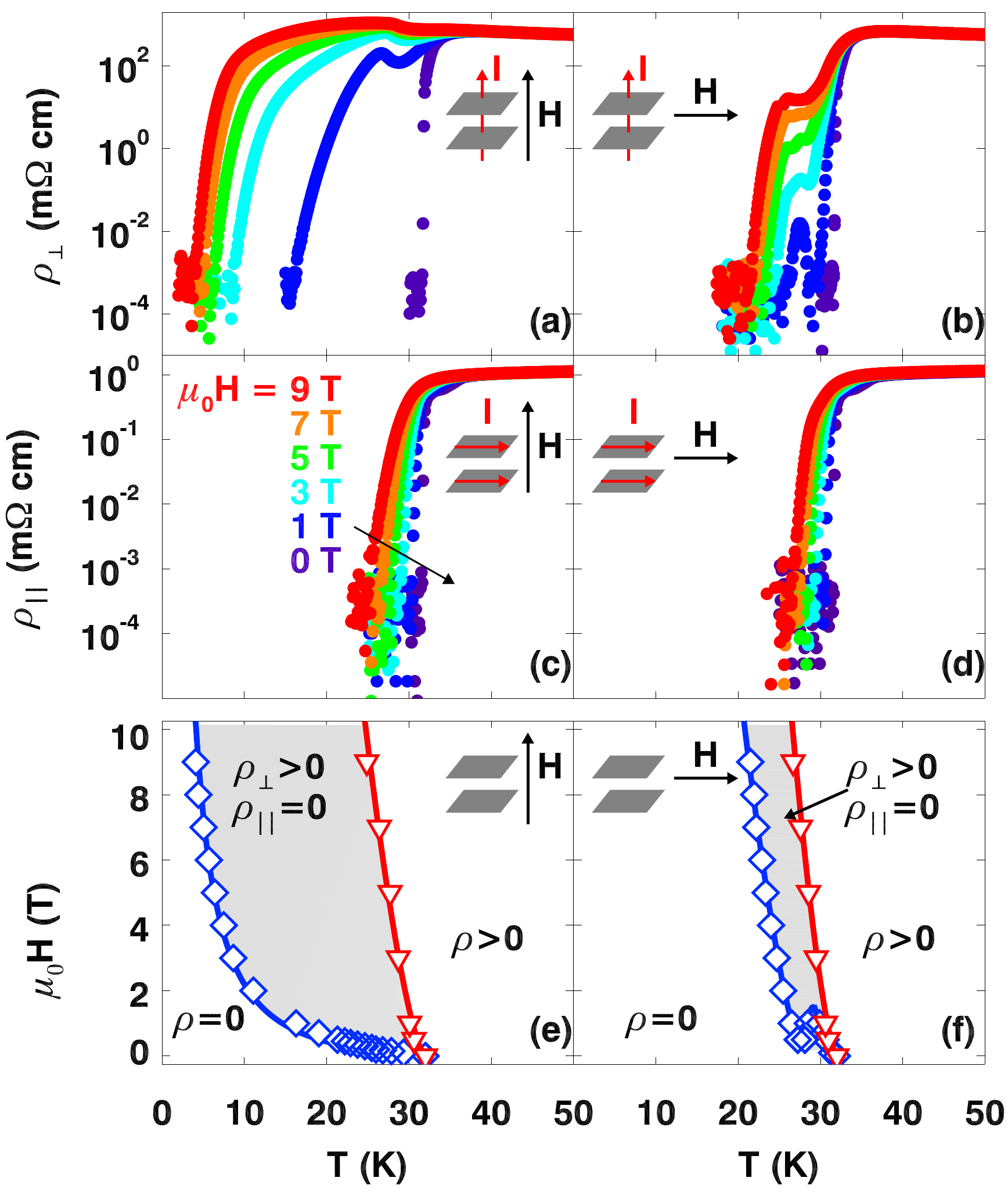}}
\caption{(Color online)
Resistivities vs.\ temperature for a range of magnetic fields, corresponding to the configurations: (a)   $\rho_\bot$ in $H_\bot$; (b) $\rho_\bot$ in $H_\|$; (c) $\rho_\|$ in $H_\bot$; (d) $\rho_\|$ in $H_\|$.  The values of $\mu_0 H$, ranging from 9~T (red) to 0~T (violet), are indicated by values and arrow in (c).  The orientations of the measuring current, $I$, and the magnetic field are indicated in the insets. (e) Phase diagram for $H_\bot$ indicating anisotropic boundaries for the onset of finite resistivity.  (f) Similar phase diagram for $H_\|$.
\label{fg:rho_log}}
\end{figure}

The results on LSCO motivated us to investigate related behavior in \lbco\ (LBCO) with $x=0.095$ and $T_c=32$~K, a sample that we have previously shown to have weak charge and spin stripe order in zero field.\cite{huck11}  In particular, we have measured the resistivity perpendicular ($\rho_\bot$) and parallel ($\rho_\|$) to the layers for various values of $H_\bot$, as shown in Fig.~\ref{fg:rho_log}(a) and (c); for comparison, we also performed measurements in fields parallel to the layers, $H_\|$, as shown in Fig.~\ref{fg:rho_log}(b) and (d).  In zero field, $\rho_\bot$ and $\rho_\|$ head towards zero at the same temperature; however, when $H_\bot$ is applied, we find that $\rho_\bot$ drops towards zero at a much lower temperature than does $\rho_\|$, as summarized in Fig.~\ref{fg:rho_log}(e).   As these results conflict with the theoretical expectations outlined above,\cite{kosh96a,ogan06} most of this paper is devoted to describing the measurements and further tests in detail, comparing with work by other groups, and considering possible spurious effects.  We conclude that we have detected intrinsic behavior.

To provide some perspective, we point out that our result for the $\rho_\bot>0$ onset curve in the $H_\bot$-$T$ phase space is quantitatively similar to the curve for the loss of superconducting order reported by others in LSCO\cite{sasa00,gila05} and LBCO\cite{adac05} for very similar doping.\footnote{To be clear, the measurements of resistivity reported in Refs.~\onlinecite{sasa00} and \onlinecite{adac05} are nominally for $\rho_\parallel$, rather than $\rho_\bot$, 
but we expect that if they had determined $T_c(H_\bot)$ from $\rho_\bot$, they would have obtained results consistent with those of Ref.~\onlinecite{gila05}.}  
The distinctive feature here is seeing the onset of finite $\rho_\|$ delayed to much higher temperatures.  For the case of  $H_\|$, summarized in Fig.~\ref{fg:rho_log}(f), the difference between the onset curves for finite $\rho_\bot$ and $\rho_\|$ is much smaller.  In measurements on a closely related material, La$_{1.6-x}$Nd$_{0.4}$Sr$_{x}$CuO$_4$, Xiang {\it et al.}\cite{xian09} have shown that the magnetoresistance measured perpendicular to the planes has a 4-fold oscillation as ${\bf H}_\|$ is rotated in the $ab$ plane; similar behavior may be expected in our case, though we have not tested for it.  

Returning to the results in $H_\bot$, which are the main focus of this paper, voltage vs.\ current measurements indicate linear resistivity at the onset of $\rho_\bot$ with nonlinear behavior at higher currents.  These observations are consistent with a model\cite{ambe69} of thermally-induced voltage fluctuations at currents below the interlayer Josephson critical current that has been applied in previous studies of $\rho_\bot$ in cuprates.\cite{bric91,gray93,hett95,suzu98b}  In contrast, $\rho_\|$ is found to be nonlinear in current, consistent with superconducting order, despite the linear $\rho_\bot$.

A complication in our sample is that there is a structural transition that overlaps with the superconducting transition.  In LBCO, there is a transition from the low-temperature orthorhombic (LTO) phase, also found in LSCO, to the low-temperature tetragonal (LTT) or low-temperature less-orthorhombic (LTLO) phase.\cite{axe94,huck11}  For the present composition, the transition is to the LTLO phase\cite{huck11}; however, it is a first-order transition, with coexisting phases over a range of temperatures.  We have characterized this behavior with neutron diffraction, thermal conductivity, and thermopower in the following paper,\cite{wen12a}  which we will refer to as paper II.  A consequence of the transition is a reduction of the interlayer Josephson coupling in the LTLO phase,\cite{home12} resulting in the sharp structure in $\rho_\bot$ near 27~K when modest $H_\bot$ is applied,\cite{wen12a} as apparent in Fig.~\ref{fg:rho_log}(a) and (b).

An interesting feature of the LTLO and LTT phases is that they are able to pin charge and spin stripe orders.\cite{tran95a,fuji04,huck11}  Here we are interested in the potential for field-induced stripe order\cite{lake02,chan08,fuji12} and possible connections with the field-induced decoupling of the superconducting layers.  
We present neutron and x-ray diffraction measurements of spin and charge stripe orders, respectively, demonstrating that both are enhanced by $H_\bot$.  To our knowledge, this is the first time that magnetic-field-induced enhancement of charge-stripe superlattice intensity has been reported.

The apparent decoupling of superconducting layers, together with the presence of stripe order, has similarities to the quasi-two-dimensional (2D) superconductivity found in LBCO with $x=1/8$.\cite{li07,tran08}  In the latter case, the quasi-2D superconductivity survived in finite $H_\bot$, though the onset temperature decreased with $H_\bot$ much more rapidly than in the present case.  While the field-dependence was not explained, a proposed explanation for the zero-field decoupling involves an intertwining between the superconductivity and stripe order.\cite{hime02,berg07}  The same type of intertwining might be involved in the field-induced state.


The rest of the paper is organized as follows.  The experimental methods, including comparisons with previous work, are discussed the next section.  The resistivity data and voltage vs.\ current measurements are discussed in Sec.~\ref{sc:resist}, while the diffraction results are presented in Sec.~\ref{sc:diff}.  The results are summarized and their implications are discussed in Sec.~\ref{sc:sum}.

\section{Experimental Methods}

\subsection{Resistivity Measurements}
\label{sc:emr}

\begin{figure}[b]
\centerline{\includegraphics[width=3.3in]{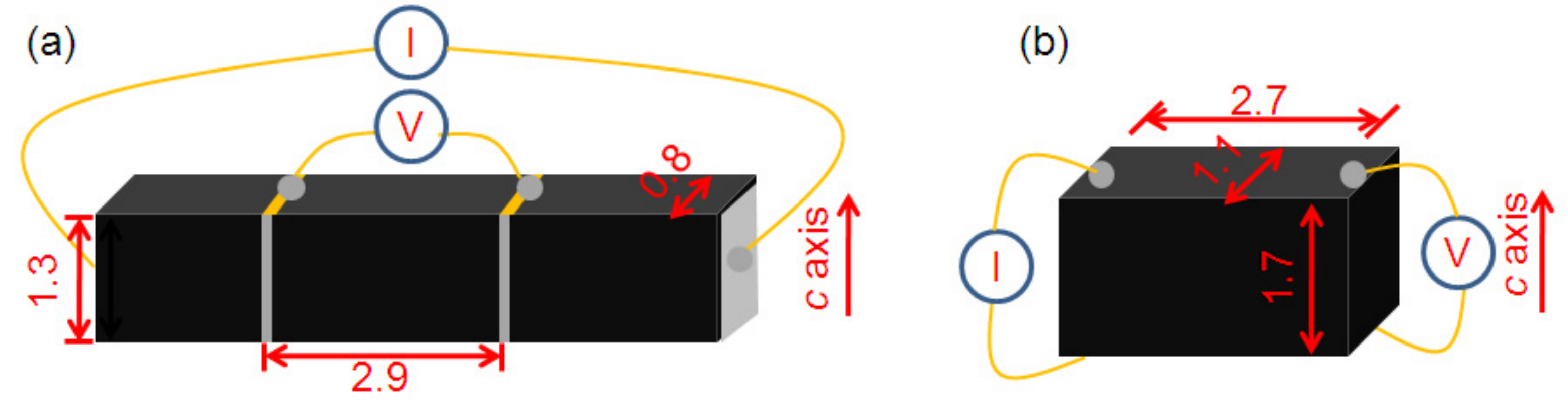}}
\caption{(Color online)  Diagrams indicating contact configurations for the measurements of (a) $\rho_\|$ and (b) $\rho_\bot$.  In (a), the voltage contacts extend across one side of the crystal to ensure adequate sensing of transport in the CuO$_2$ planes.
\label{fg:contacts}}
\end{figure}

The crystals for this study were grown by the traveling-solvent, floating-zone technique; characterizations of the crystals are reported in paper II and in Refs.~\onlinecite{huck11,home12}.  The resistivity measurements were performed at Brookhaven by the standard four-probe technique in a Physical Properties Measurement System (Quantum Design).  Different crystals, cut from the same parent, were used for the $\rho_\bot$ and $\rho_\|$ measurements, and the $\rho_\|$ results were confirmed on a third crystal.  For $\rho_\bot$, the crystal dimensions ($l \times w \times t$) were $1.7\times 2.7 \times 1.1$~mm$^3$ with voltage and current contacts on the $a$-$b$ faces;  for $\rho_\|$, the dimensions were $8.0\times 0.8 \times 1.3$~mm$^3$ with 2.9 mm between voltage contacts.  Contacts, made with Ag paste, were annealed at $400^{\circ}$C for 1~h; the contact configurations are illustrated in Fig.~\ref{fg:contacts}.  In each case, contact resistance was measured at room temperature and confirmed to be less than 2~$\Omega$ before and after the transport studies.  The dc measuring current was 1 mA, corresponding to current densities of $J_\bot=0.03$~A/cm$^2$ and $J_\|=0.1$~A/cm$^2$, and repeated measurements at each temperature were averaged.  In the voltage vs.\ current measurements, the dc current was varied from 10 nA to 5 mA.

\begin{figure}[t]
\centerline{\includegraphics[width=3.3in]{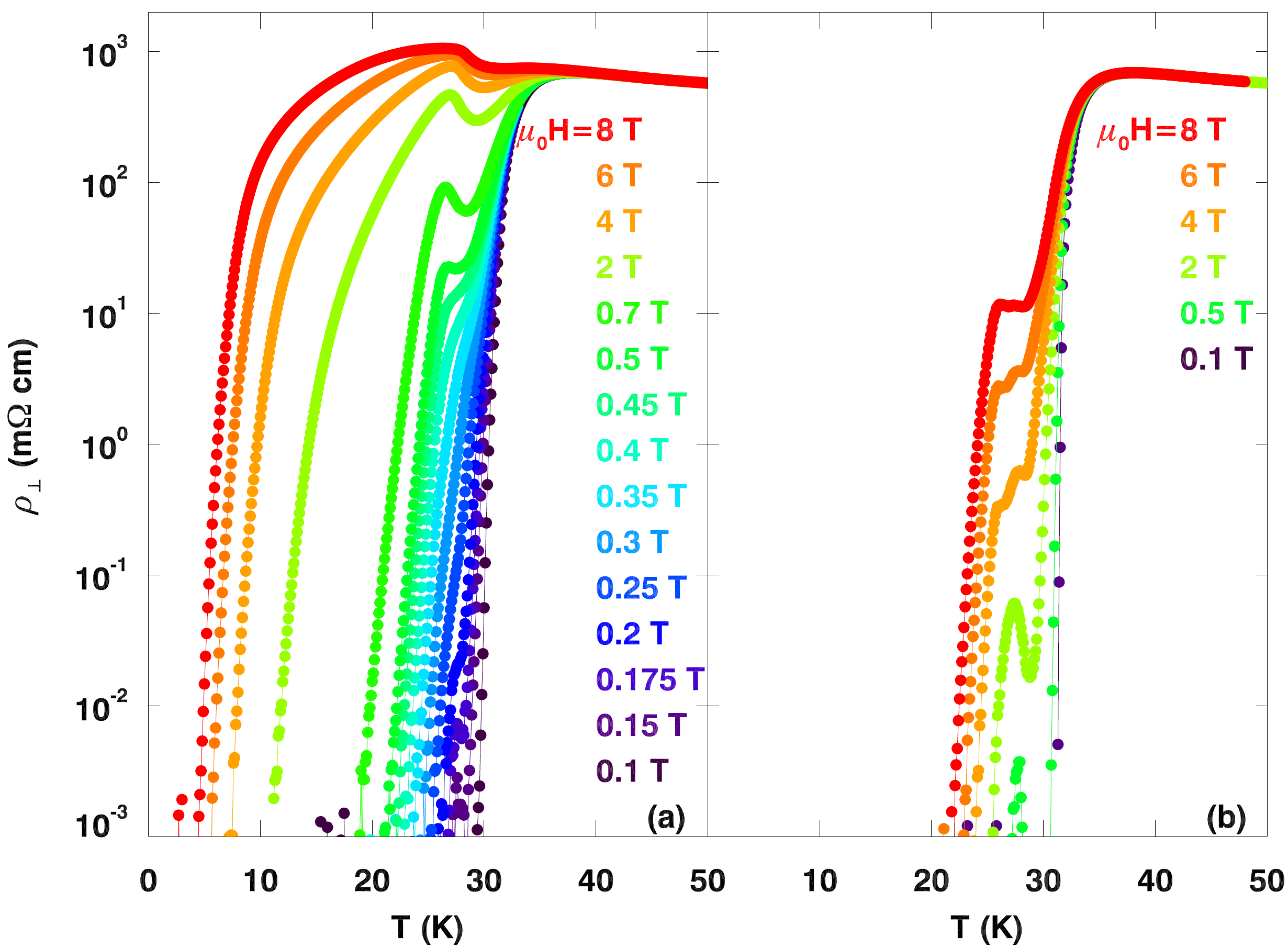}}
\caption{(Color online)  
Further data sets for $\rho_\bot$ with: (a) $H_\bot$, (b) $H_\|$.  Magnetic field values are listed in the legends.
\label{fg:rho_more}}
\end{figure}

To define the boundaries between the regions of negligible and finite resistivity in Fig.~\ref{fg:rho_log}(e) and (f), we used a finite resistivity threshold of $1\times 10^{-3}$ m$\Omega$ cm.  Measured values below that level tend to fluctuate around zero (or rather the measured voltage fluctuates about zero); in Fig.~\ref{fg:rho_log} and corresponding figures we have actually plotted the absolute value of the resistivity so that all points appear on the logarithmic scale.   We will discuss the behavior of $\rho_\|$ in $H_\bot$ near the threshold further below.   To complete the phase boundaries for $\rho_\bot$, especially at lower fields, we made additional measurements, which are shown in Fig.~\ref{fg:rho_more}.

\begin{figure}[b]
\centerline{\includegraphics[width=3.3in]{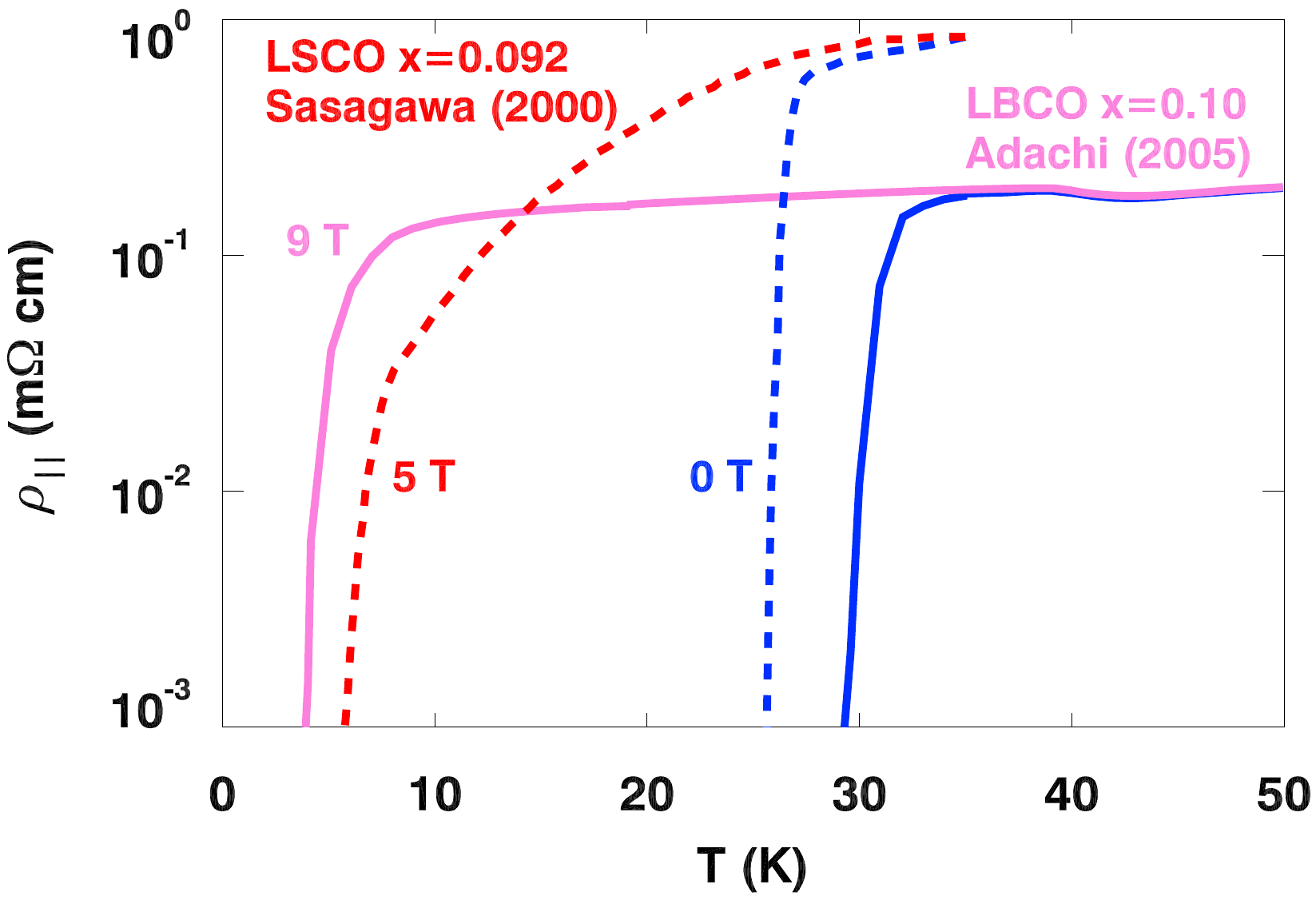}}
\caption{(Color online)  Measurements of $\rho_\|$ in LSCO with $x=0.092$ from Sasagawa {\it et al.}\cite{sasa00}  (dashed lines) and in LBCO with $x=0.10$ from Adachi {\it et al.}\cite{adac05} (solid lines), with values of $\mu_0H_\bot$ labeling the curves.
\label{fg:compare}}
\end{figure}

Given the unusual nature of our observations for $\rho_\|$ and $\rho_\bot$ in $H_\bot$, it is worthwhile to compare with measurements by other groups on similar samples.  In Fig.~\ref{fg:compare}, we plot results of $\rho_\|$ for LSCO with $x=0.092$ reported by Sasagawa {\it et al.}\cite{sasa00} in $\mu_0H_\bot$ of 0 and 5~T, and results for LBCO with $x=0.10$ from Adachi {\it et al.}\cite{adac05} in perpendicular fields of 0 and 9~T.  The latter sample exhibits an LTO-LTT structural transition on cooling through 40~K, causing a small step in the resistivity.  That transition is at a higher temperature than in our $x=0.095$ crystal, consistent with a difference in Ba concentration.\cite{huck11}  As one can see, the impact of the field on $\rho_\|$ is very similar to what we find for $\rho_\bot$ in Fig.~\ref{fg:rho_log}(a), but it is grossly different from what we find for $\rho_\|$ with our sample, as shown in Fig.~\ref{fg:rho_log}(c). 

It has been reported that there can be complications in interpreting resistivity measurements in the mixed phase.\cite{fuch98}  When measuring $\rho_\|$ in $H_\bot$ above the lower critical field, there is always a Lorentz force on the vortices.  If the vortices are not pinned, the Lorentz force results in vortex motion and dissipation.  In a study by Fuchs {\it et al.},\cite{fuch98} it was demonstrated in a crystal of \bscco\ that a surface barrier effect can inhibit vortex motion near the surface, while vortices can flow within the bulk.  As a consequence of this inhomogeneous flux pinning, an applied current tends to flow primarily along the crystal edges, even when the flux-flow resistivity is rather high, such as when $T$ approaches $T_c$.  This work was extended more recently by Beidenkopf {\it et al.}\cite{beid09}  

One complication in the latter work is that the current and voltage contacts were all located on $a$-$b$ surfaces, so that the current had to flow along the $c$-axis in order to reach the interior CuO$_2$ planes.  (This effect is explained by an earlier analysis of Busch {\it et al.}\cite{busc92})  The researchers found it necessary to irradiate the current contacts with a beam of high-energy Pb ions in order to create vortex pinning centers, and thus inhibit flux-flow resistivity along the $c$ axis; this significantly reduced the effective resistance along the in-plane direction in the mixed state.\cite{beid09}  In any case, we will discuss the possible impact of inhomogeneous current flow associated with a flux-pinning surface barrier in Sec.~\ref{sc:rho_par}.

\begin{figure}[b]
\centerline{\includegraphics[width=3.3in]{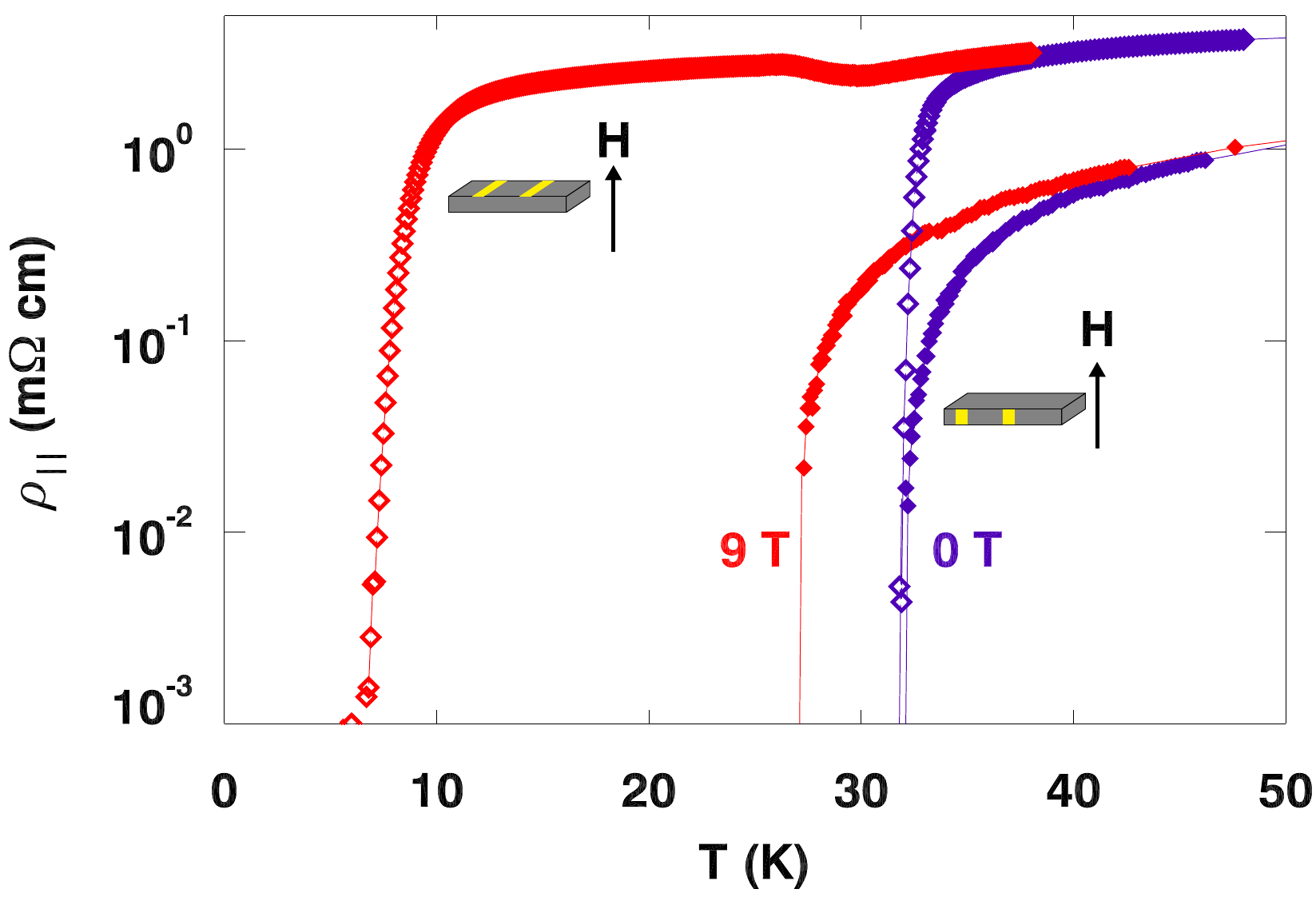}}
\caption{(Color online)  
Results for $\rho_\|$ measured in $H_\bot$ with voltage contacts on a crystal face perpendicular to the crystallographic $c$ axis (open diamonds)  and parallel to $c$ (filled symbols), as indicated by the insets.  In both cases, violet symbols correspond to zero field, red to $\mu_0 H_\bot=9$~T.
\label{fg:bad}}
\end{figure}


In our studies of LBCO, we have observed the impact of slightly different contact configurations. We use current contacts covering opposite crystal faces, both of which are normal to the $a$-$b$ planes.  In Fig.~\ref{fg:bad}, we compare attempts to measure $\rho_\|$ with voltage contacts A)  on an $a$-$b$ face, and B) on a side face normal to the $a$-$b$ planes [consistent, in effect, with Fig.~\ref{fg:contacts}(a)].  The two measurements give a consistent result for $T_c$ in zero field; however,  in $\mu_0H_\bot=9$~T, configuration A shows a signal that appears to mimic $\rho_\bot$ in temperature dependence below the zero-field $T_c$, while configuration B indicates a sharp drop in $\rho_\|$ at a much higher temperature.  (We observed identical sensitivity to contact configuration in studying LBCO $x=1/8$, although we did not report on it there.\cite{li07})  We interpret the results as follows.  The direction of low-resistivity is always parallel to the planes, so we expect the current flow to be uniform across all planes that are in contact at both ends (except in the the possible case of flux-flow conditions and a vortex-pinning surface barrier, as discussed above).  The sample is sufficiently thick and well-oriented that most planes satisfy this condition.  It follows that the contacts in configuration B will sense the voltage drop due to transport parallel to the planes alone, thus directly probing $\rho_\|$.  In the case of configuration A, inevitable imperfections in sample orientation mean that the voltage contacts on the top surface can only sense the transport if there is some flow perpendicular to the planes, so that the effective resistivity has a contribution from $\rho_\bot$.  

We are certainly not the first to make measurements with contacts on crystal sides that are perpendicular to the $a$-$b$ planes.  For example, Cho {\it et al.}\cite{cho94} used contact configurations similar to ours in their magnetic transport study of \bscco.  Their data appear to show that $\rho_\|$ heads to zero at a higher temperature than $\rho_\bot$ for finite $H_\bot$, although they did not comment on this point.   

Let us return for a moment to the earlier data on LSCO and LBCO shown in Fig.~\ref{fg:compare}.  It is possible that the response in LSCO is different from what we observe in LBCO, as LSCO retains the LTO structure down to low temperature.  We will argue later that the LTLO structure could be relevant to the unusual behavior we observe for $\rho_\|$.  On the other hand, the LBCO $x=0.10$ sample studied by Adachi {\it et al.}\cite{adac05} should exhibit extremely similar behavior to our $x=0.095$ crystals.  We note that their measurements of $\rho_\|$ look quite similar to the results in Fig.~\ref{fg:bad} for the configuration that is limited by a contribution from $\rho_\bot$; however, the details of the contact configuration are not given in their paper.\cite{adac05}

\subsection{Scattering Measurements}

The neutron diffraction measurements were performed on the SPINS spectrometer at the NIST Center for Neutron Research, with an incident energy of 5~meV and a cooled Be filter after the sample to minimize intensity at harmonics of the desired neutron wave length. Horizontal collimations were $55'$-$80'$-S-$80'$-$240'$.  The sample (S) was a cylindrical crystal with a mass of 11.2~g, mounted in a superconducting, vertical-field magnet, allowing scattering at wave vectors $(h,k,0)$, in reciprocal lattice units $a^*=2\pi/a$, with $a=3.79$~\AA\ corresponding to the high-temperature tetragonal phase.  The x-ray diffraction measurements were performed at beam line BW5 at DESY using 100 keV photons ($\lambda=0.124$~\AA).\cite{huck11}  The sample was a disk 5~mm in diameter and 1 mm in thickness, oriented such that the charge-order reflection was measured in transmission geometry.

\begin{figure}[t]
\centerline{\includegraphics[width=3.in]{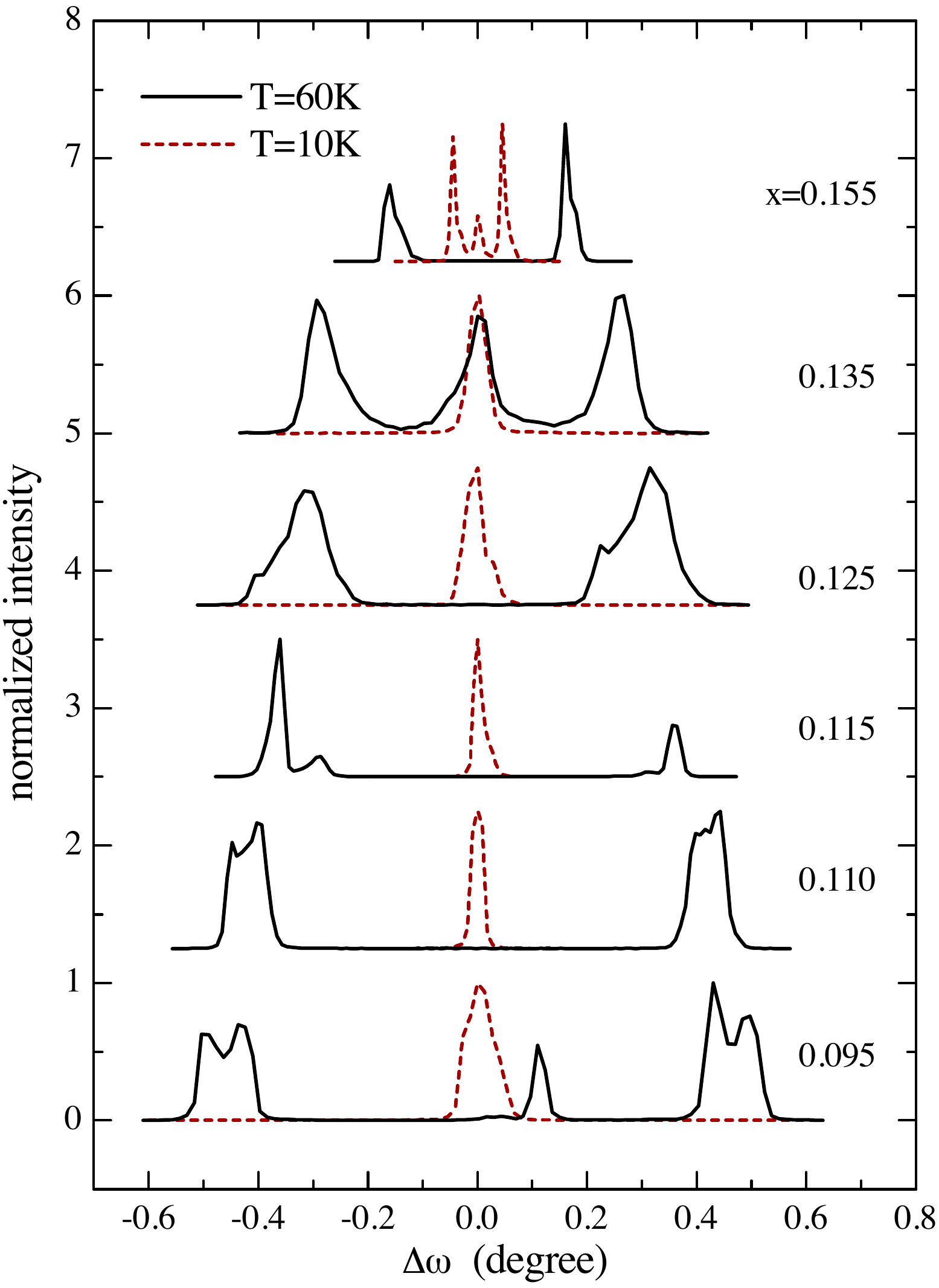}}
\caption{(Color online)  Rocking curve scans at the (200)/(020) Bragg reflection for \lbco\ crystals with a range of Ba concentrations, as indicated.   Solid lines correspond to scans at $60\pm10$~K, above the structural transition; dashed lines were measured at approximately 10~K, below the transition.  The peaks split symmetrically about zero at 60~K correspond to twinned reflections. For $x=0.095$ at 60~K, the extra peak at $0.1^\circ$ corresponds to a small-volume domain with a more complicated twin relationship.
\label{fg:ortho}}
\end{figure}

The x-ray measurements provide a useful measure of the degree of uniformity of the crystals.  Figure~\ref{fg:ortho} shows rocking curves through the (200) reflection for a series of LBCO crystals with a range of Ba concentrations.  (We recently reported\cite{huck11} the phase diagram for this system based on diffraction data.)  The measurements at ``60 K'' are $\pm10$~K; they are in the LTO phase above the structural transition to the LTT (or LTLO) phase.  For a perfect, single-domain crystal there would be a single peak.  Because of twinning, the peaks in our crystals are split about zero.  Note that the magnitude of the splitting is determined by the orthorhombic strain, which decreases as $x$ grows.  There is also a finer scale splitting, that appears to be from a small monoclinic distortion.\cite{reeh06}  At base temperature, $\sim10$~K, there is a single peak for the LTT phase; in the case of $x=0.095$, there is a small peak splitting (unresolved here) due to the small orthorhombic distortion of the LTLO phase.

The message here is that, based on diffraction evidence, each crystal has a rather uniform composition that is narrowly defined.  Keep in mind that the x-ray beam samples the full 1-mm thickness of each crystal.  The compositions are also distinguished by the structural transition temperatures, as discussed in Ref.~\onlinecite{huck11}. 

\section{Resistivity Results}
\label{sc:resist}

\subsection{Overview}

\begin{figure}[b]
\centerline{\includegraphics[width=3.3in]{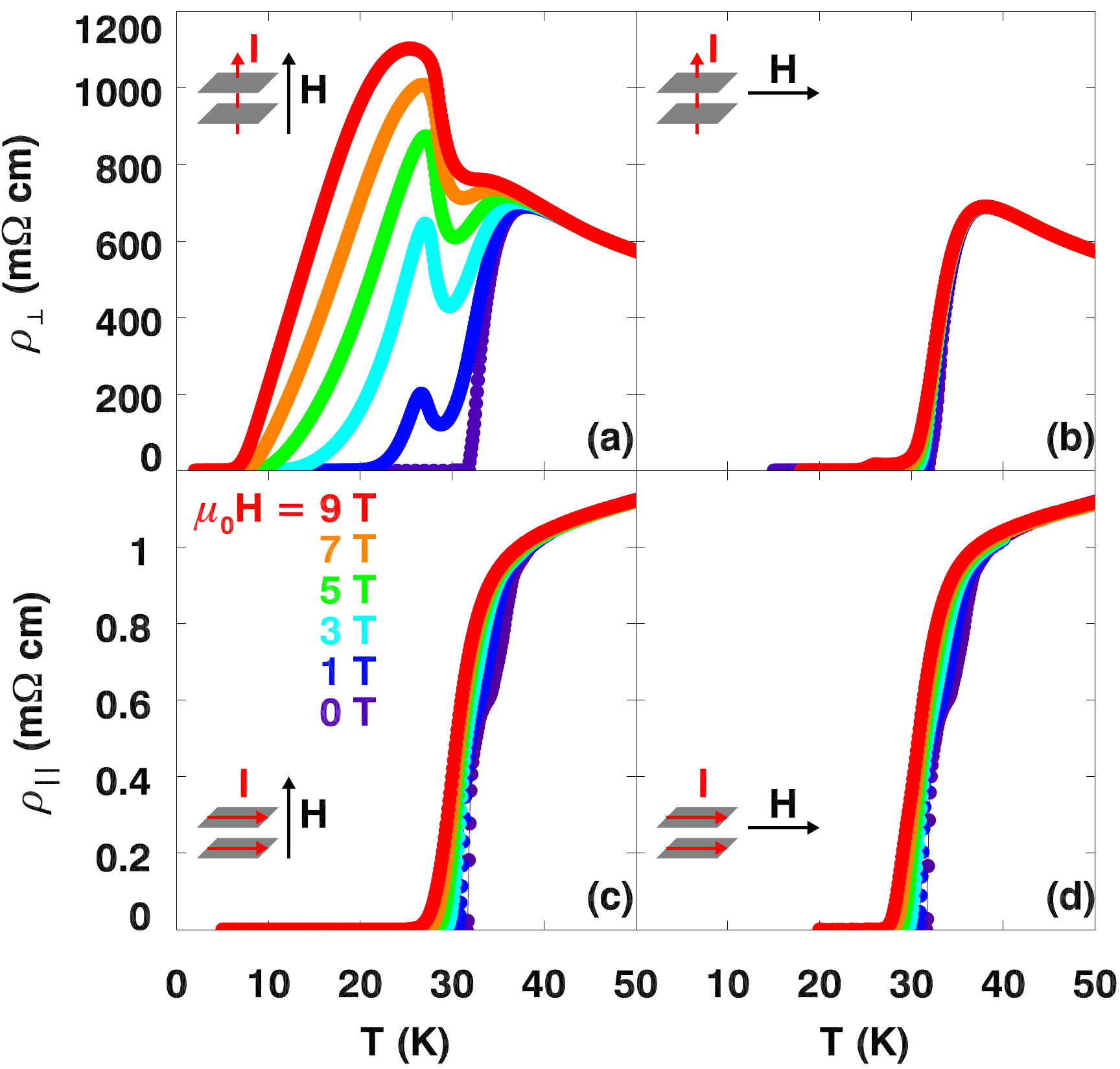}}
\caption{(Color online)  
Resistivities vs.\ temperature for a range of magnetic fields, corresponding to the configurations: (a)  $\rho_\bot$ in $H_\bot$, (b) $\rho_\bot$ in $H_\|$, (c) $\rho_\|$ in $H_\bot$, (d) $\rho_\|$ in $H_\|$.  The values of $\mu_0 H$, ranging from  9~T (red) to 0~T (violet), are indicated in (c).  The orientations of the measuring current, $I$, and the magnetic field are indicated in the insets.
\label{fg:rho_lin}}
\end{figure}

To provide an alternative view of the resistivity data, we have plotted the results with linear scales in Fig.~\ref{fg:rho_lin}.   This form makes it easier to see that $\rho_\bot$ continuously increases with $H_\bot$ for $T\alt27$~K.  In other words, we are not able to apply a large enough $H_\bot$ to cause $\rho_\bot$ to reach its normal state behavior for $T\alt27$~K.  As demonstrated in detail in paper II, 27~K corresponds to the completion of the LTO to LTLO structural transition.  Associated with the transition is a reduction in the interlayer Josephson coupling, as discussed in II; further support for this effect is provided by a study of the temperature dependence of the $c$-axis Josephson plasma resonance.\cite{home12}  We have argued in II that it is the change in the interlayer Josephson coupling associated with the transition that causes the structure in $\rho_\bot(T)$ when measured in $\mu_0H_\bot$ as low as 0.15~T.  There is also a step in $\rho_\|$ at $\sim35$~K measured in zero field; we have shown in paper II that this corresponds to the onset of the structural transition. 

\subsection{Analysis of $\rho_\|$}
\label{sc:rho_par}

With unusual behavior, it is important to consider possible extrinsic explanations.  While the coincidence of the structural and superconducting transitions complicates the situation, it does not provide an explanation for the regime of apparent uniaxial resistivity.  Could there be some sort of inhomogeneity in the samples that causes the $\rho_\|$ and $\rho_\bot$ measurements to be determined by distinct phases?  Before addressing this question, we point out that the phase boundary for superconducting order determined by our $\rho_\bot$ data is quantitatively consistent with results in the literature\cite{sasa98,ando99b,sasa00,lake02,adac05,gila05} for LSCO and LBCO with $x\sim0.1$.  It follows that only the behavior of $\rho_\|$ is anomalous.  So, could there be layered intergrowths of a more robust superconducting phase that might only be detected in the $\rho_\|$ configuration?  We see no credible way for this to occur.  First of all,  the sample studied here (LBCO $x=0.095$) has the highest zero-field $T_c$ in the LBCO phase diagram,\cite{huck11} so that compositional variation could only lead to regions of reduced $T_c$, and that would not explain our observations.  Secondly, we have presented diffraction data in Fig.~\ref{fg:ortho}, in paper II, and elsewhere,\cite{huck11} as well as thermodynamic measurements in paper II, that indicate high quality samples with no indications of unique compositional inhomogeneities.  Finally, if there were special layers present of a superconductor capable of providing negligible $\rho_\|$ in substantial $H_\bot$, then one might expect these layers to have a higher zero-field $T_c$ than that detected in $\rho_\bot$, but there is no evidence of such an inhomogeneous anisotropy.

Another possibility to consider is inhomogeneous current flow below $T_c$ in the mixed phase.  As discussed in Sec.~\ref{sc:emr}, studies of \bscco\ have shown that in the regime where vortices are not uniformly pinned, the in-plane current tends to flow along the edges of the crystal.\cite{fuch98,beid09}  For our contact configuration, such an effect would increase the current density in the vicinity of the voltage contacts, with the consequence that $\rho_\|$ would be underestimated.  If $\rho_\|$ were driven below our threshold sensitivity, then we would overestimate the temperature at which $\rho_\|$ becomes negligible.  Thus, this effect could cause a quantitative error in our analysis.  The bigger question, though, is whether this effect could explain the difference between our measurements of $\rho_\|$ in large $H_\|$ and that reported by other groups on similar samples\cite{sasa00,adac05} (see Fig.~\ref{fg:compare}).  The straightforward answer is that it cannot. The latter data suggest that the normal state extends down to temperatures well below the point where we observe a drop due to superconducting correlations.

We have suggested that the differences for $\rho_\|$ between our results and previous work might be due to different contact configurations.  As this can be a contentious issue, it is highly desirable to confirm our results with independent measurements.  We present two such checks below.

\begin{figure}[t]
\centerline{\includegraphics[width=3.1in]{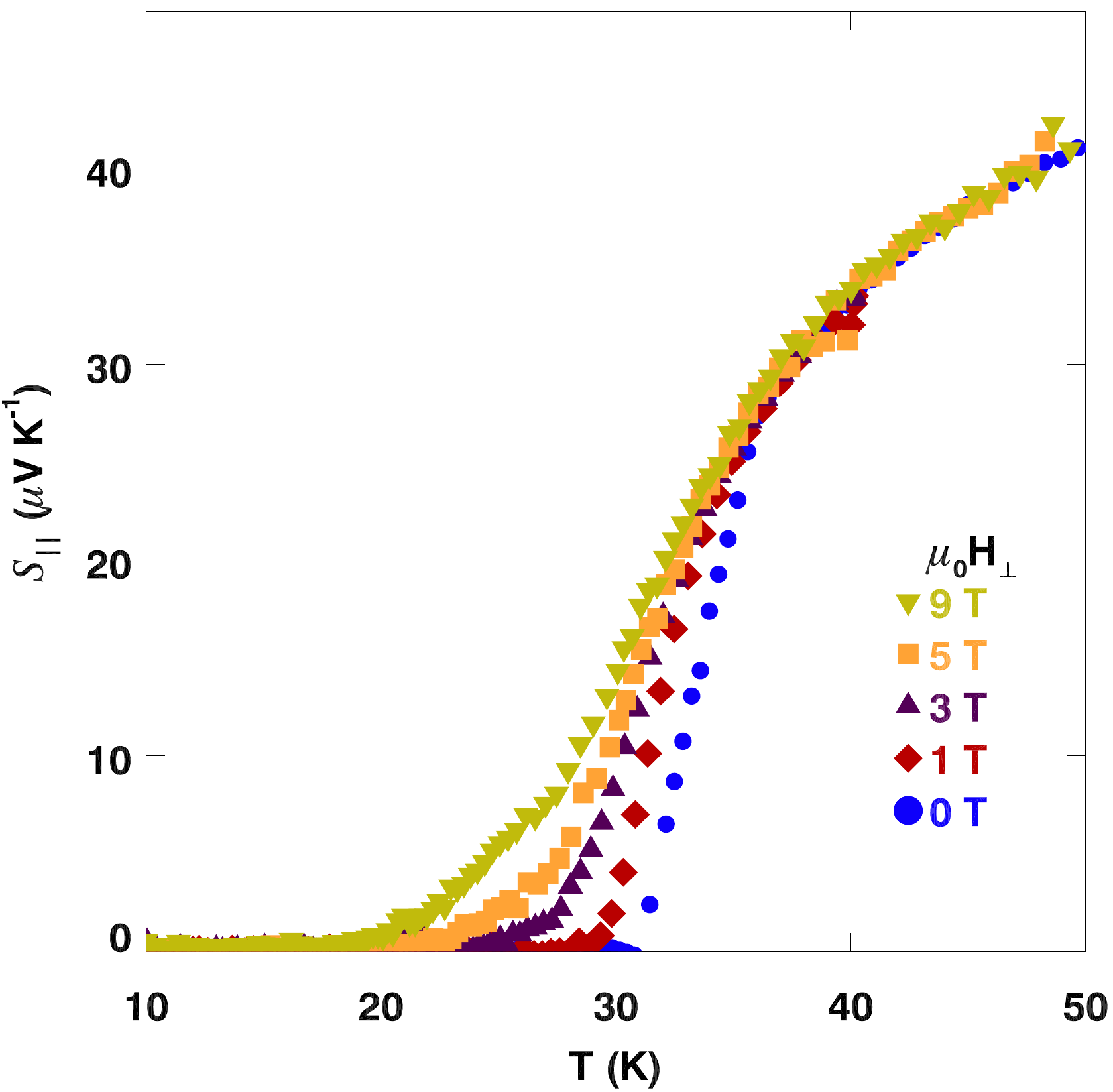}}
\caption{(Color online)
Measurements of the in-plane thermopower, $S_\|$, with $\mu_0H_\bot=0$, 1, 3, 5, and 9~T.
\label{fg:S}}
\end{figure}

One confirmation is provided by field-dependent measurements of the in-plane thermopower, $S_\|$, shown in Fig.~\ref{fg:S} (experimental details and further discussion are presented in paper II).   Note that the thermopower must be zero in the superconducting state.   As one can see, $S_\|$ drops towards zero on cooling in high field in a fashion quite similar to $\rho_\|$ and clearly distinct from $\rho_\bot$.  From these data, one might arrive at a slightly lower estimate for the temperature at which in-plane superconducting order appears at high field, but it would still be well above the point at which $\rho_\bot$ heads to zero.

\begin{figure}[t]
\centerline{\includegraphics[width=3.3in]{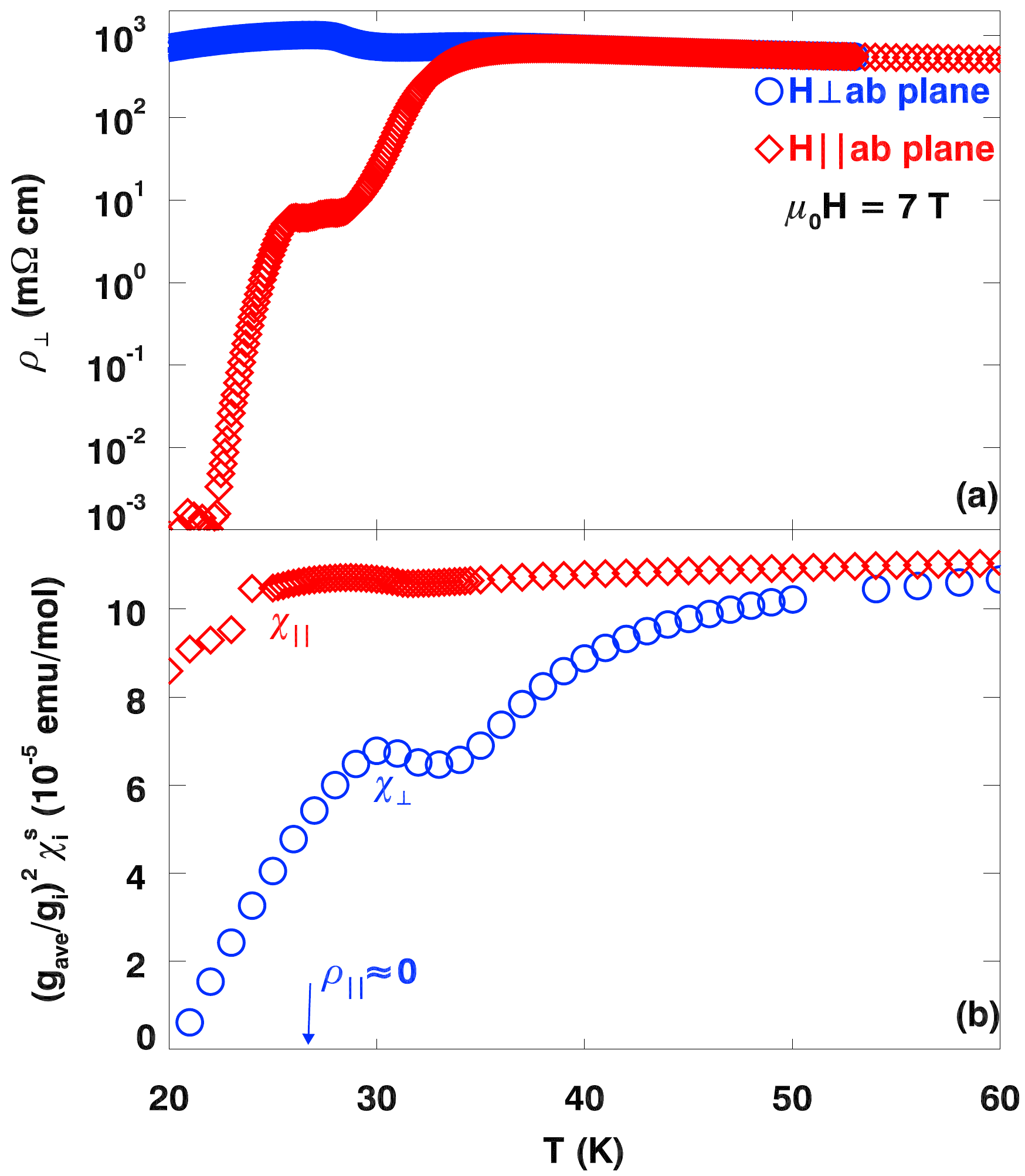}}
\caption{(Color online)
(a) Data for $\rho_\bot$ in $\mu_0H=7$~T applied both parallel (diamonds) and perpendicular (circles) to the planes.  (b)  Field-cooled spin susceptibility in the form $(g_{\rm ave}/g_i)^2\chi_i^s$ for $i = \|$ (diamonds) and $\bot$ (circles) obtained with $\mu_0H=7$~T, as discussed in the text.
\label{fg:susc}}
\end{figure}

A second confirmation is provided by high-field measurements of the magnetic susceptibility ($\chi=M/H$).  The data for $\mu_0H=7$~T, for fields both parallel and perpendicular to the planes, were already presented in Fig.~10 of Ref.~\onlinecite{huck11}.  In Fig.~\ref{fg:susc}(b), we present the spin susceptibilities in the form $(g_{\rm ave}^2/g_i^2)\chi_i^s$ ($i=\|$, $\bot$) after correction for Van Vleck susceptibility, core diamagnetism, and anisotropic $g$ factors, as explained in paper II and in Ref.~\onlinecite{huck08}.  The data for $\rho_\bot$ in the same fields are reproduced in Fig.~\ref{fg:susc}(a).

The significance of this comparison may require some explanation.  First consider $\chi_\bot^s$, which is indicated by the (blue) circles in Fig.~\ref{fg:susc}(b).  It shows a growth of diamagnetism ({\it i.e.}, a decrease of $\chi_\bot^s$ below the paramagnetic susceptibility of the normal state) on cooling that reflects diamagnetic screening currents within the planes.  This growth of superconductivity in the planes is completely missed by $\rho_\bot$, indicated by (blue) circles in Fig.~\ref{fg:susc}(a); however, it is  consistent with $\rho_\|(\mu_0H_\bot=7~\mbox{\rm T})$ [see Fig.~\ref{fg:rho_log}(c)].  To emphasize how extreme a situation this is, it is instructive to compare with the measurements in $H_\|$, indicated by the (red) diamonds in Fig.~\ref{fg:susc}.  We see that $\rho_\bot(H_\|)$ drops rapidly with temperature compared to $\rho_\bot(H_\bot)$.  This happens despite the fact that the diamagnetism measured by $\chi_\|$ does not become apparent before the temperature reaches $\sim22$~K.  Here, the diamagnetic response requires screening currents that loop between the layers.  The drop in $\rho_\bot(H_\|)$ reflects the superconducting order in the planes indicated by $\rho_\|(H_\|)$, shown in Fig.~\ref{fg:rho_log}(d).  It is due to the transport of pairs between the layers, but there is no diamagnetic screening of fields between the layers because of the lack of interlayer coherence.  Thus, the measurements of diamagnetism in high fields support the interpretation of decoupled superconducting layers indicated by the resistivity data.

\begin{figure}[t]
\centerline{\includegraphics[width=3.2in]{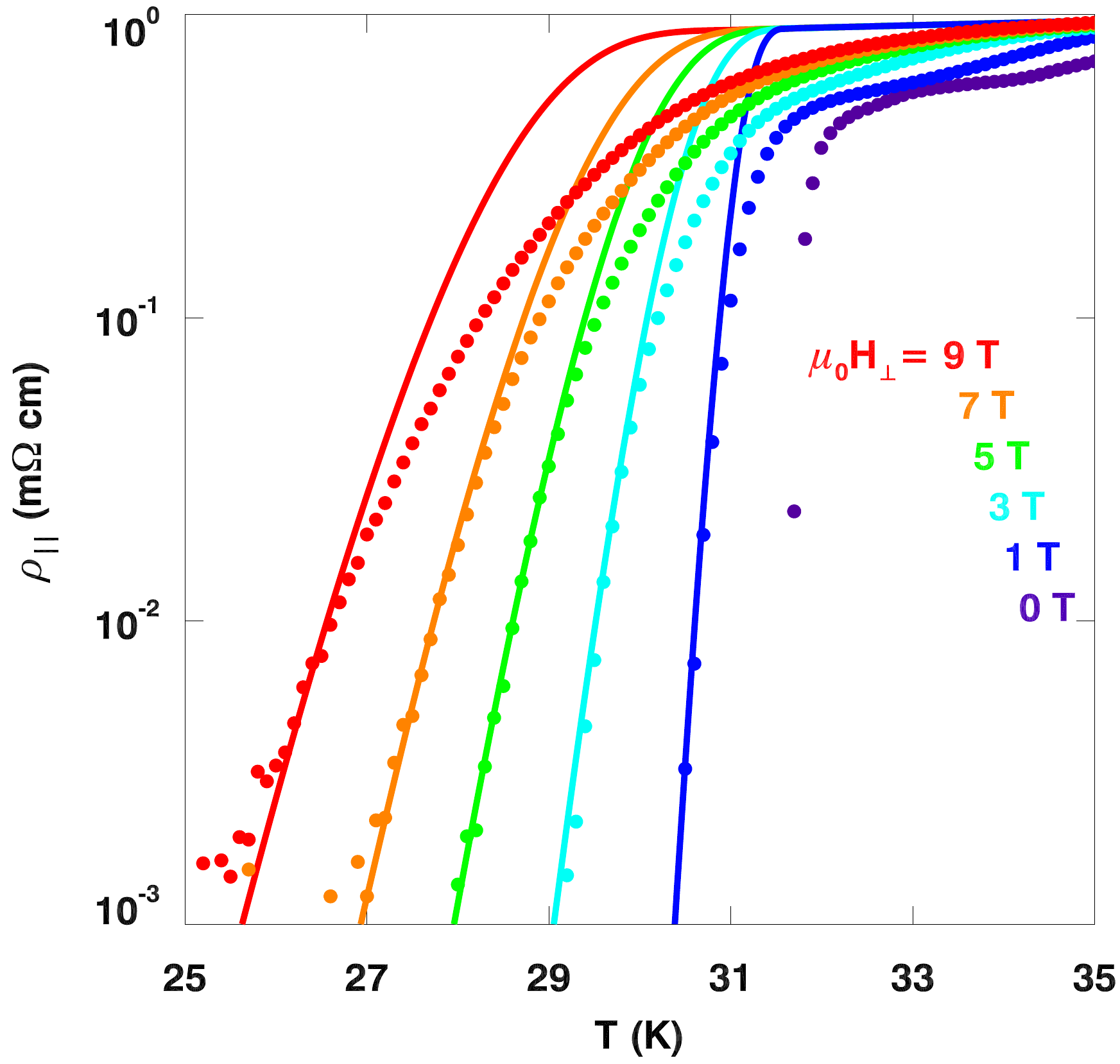}}
\caption{(Color online)  Points indicate measured $\rho_\|$ vs.\ $T$ for various values of $\mu_0H_\bot$, as indicated in the legend.  Lines are fitted curves corresponding to a model of flux-flow resistivity,\cite{tink88} as discussed in the text.
\label{fg:flux_flow}}
\end{figure}

Now, let us return to the flux-flow issue, which is expected to limit $\rho_\|$ (but not $\rho_\bot$) close to $T_c$ when $H_\bot$ is applied.  Tinkham\cite{tink88} argued that the resistivity in this case should be described by the formula
\begin{equation}
 \rho_\| / \rho_{\|n} = [I_0(\gamma_0/2)]^{-2},
\end{equation}
where $\rho_{\|n}$ is the normal state resistivity, $I_0$ is the modified Bessel function, and
\begin{equation}
 \gamma_0 = A [1-T/T_c(H)]^{3/2}/B.
\end{equation}
Here, $A\approx 0.032\times J_{\bot c0}/T_c$ is a constant having units of T (provided that $T_c$ is measured in K),  with $J_{\bot c0}$ (measured in A/cm$^2$) being the critical current density at zero temperature and zero field along the direction in which the field is applied; we take $B\approx\mu_0H_\bot$.  A fit of this formula to our $\rho_\|$ data is shown in Fig.~\ref{fg:flux_flow}.  As one can see, it gives a reasonable description of the data where $\rho_\|$ is small, and does a good job capturing the field-dependence. (In fact, it is a surprisingly good description given that we believe that the Josephson coupling is changing in an anomalous fashion in this temperature range due to the underlying structural transition.\cite{wen12a}) In the fit, $T_c$ varies from 31.6 K at 1 T to 30.9 K at 9 T.  From the fitted value of $A=1.3\times10^3$~T, we obtain the estimate $J_{\bot c0} \sim 1.3\times 10^6$ A/cm$^2$, which is of the same magnitude as that found by Tinkham\cite{tink88} in fitting similar data for \ybco\ with $T_c = 91$~K.

\begin{figure}[b]
\centerline{\includegraphics[width=3.3in]{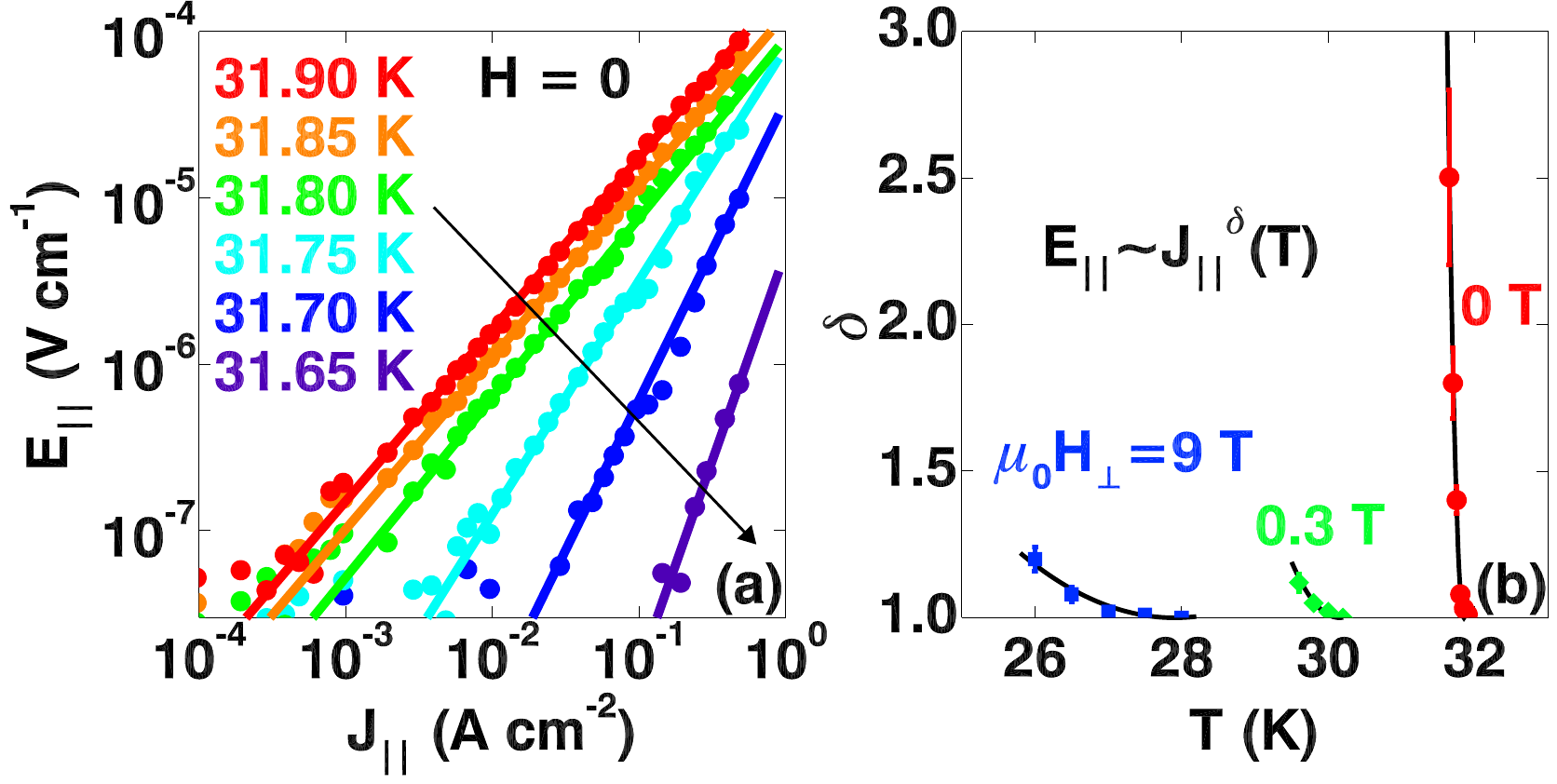}}
\caption{ (Color online)
(a) $E_\|$ vs.\ $J_\|$ in zero field for $T\sim T_c$.  (b) Exponent from power-law dependence of $E_\|$ vs.\ $J_\|$ for several values of $H_\|$.  Error bars, corresponding to one standard deviation, were determined from least squares fits to the data. 
\label{fg:test2}}
\end{figure}

As another test of the in-plane response, we have measured voltage vs.\ current behavior close to $T_c$.  An example of the measurements is shown in Fig.~\ref{fg:test2}(a), where we use the intensive quantities, in-plane electric field, $E_\|$, and in-plane current density, $J_\|$.  We observe that $E_\| \sim J_\|^\delta$, where $\delta=1$ in the normal state, while $\delta$ rapidly grows on cooling through $T_c$.  We have repeated these measurements in fields of $\mu_0 H_\bot = 0.3$ and 9 T; the evolution of the exponent $\delta$ is shown in Fig.~\ref{fg:test2}(b).    The nonlinear behavior is consistent with the onset of superconducting order associated with a vortex glass state.\cite{fish91}

\begin{figure}[b]
\centerline{\includegraphics[width=3.3in]{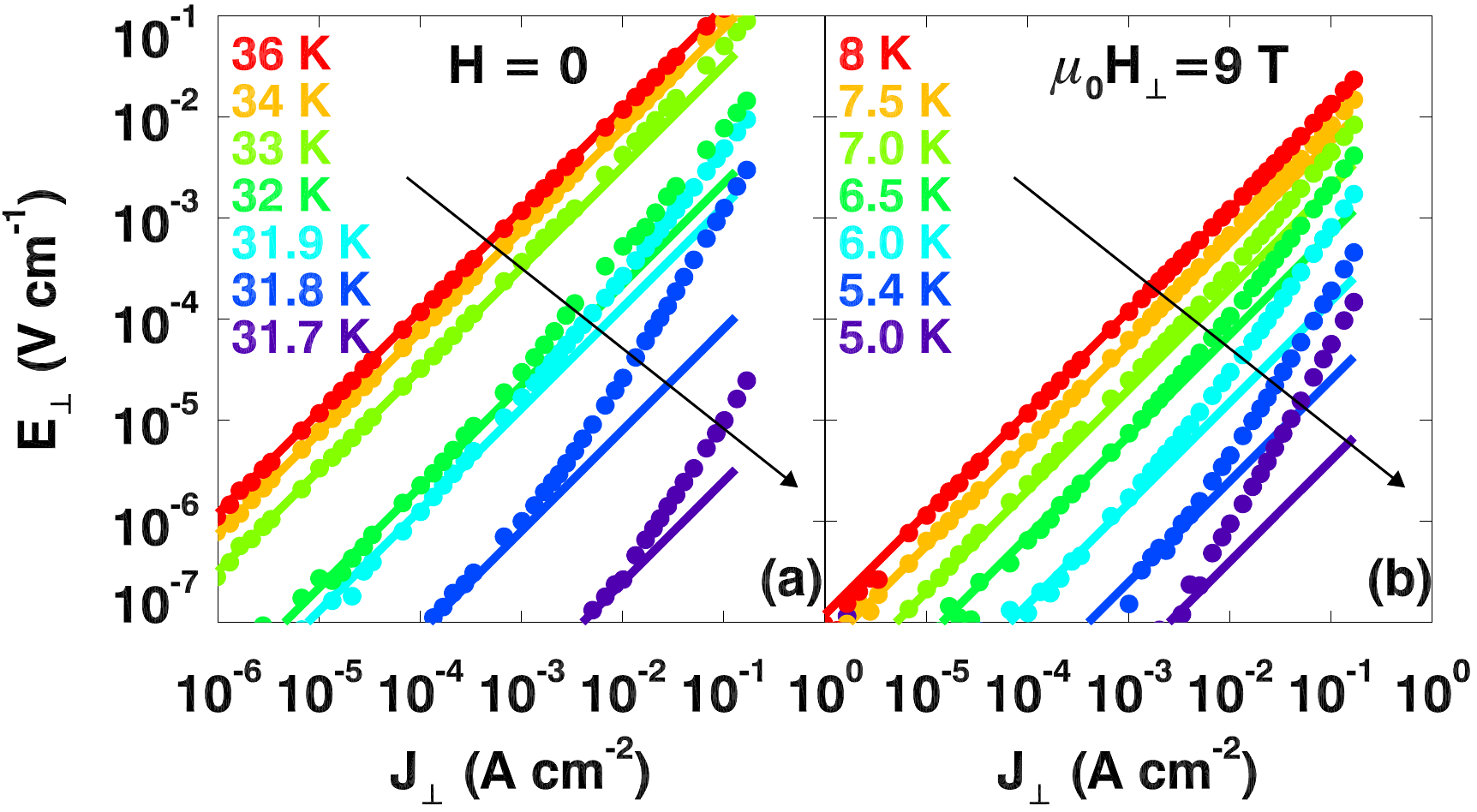}}
\caption{(Color online)
$E_\bot$  vs.\ $J_\bot$ in (a) zero field for temperatures from 36~K down to 31.7~K, as indicated in the legend, and (b) $\mu_0 H_\bot=9$~T for temperatures from 8~K down to 5~K.  Lines indicate $E_\bot\sim J_\bot$.
\label{fg:testc}}
\end{figure}

\subsection{Analysis of $\rho_\bot$}

The situation is rather different when we measure in the direction perpendicular to the planes.  As shown in Fig.~\ref{fg:testc}(a) for temperatures near $T_c$, we observe a linear relation between $E_\bot$ and $J_\bot$ at the lowest $J_\bot$ values, but a nonlinear increase in $E_\bot$ at larger $J_\bot$.  Similar behavior is found at low temperature and $\mu_0H_\bot=9$~T, as one can see in Fig.~\ref{fg:testc}(b).  The trend suggests that the linear $\rho_\bot$ extends down to negligibly small $J_\bot$.  This behavior is reminiscent of the theoretical results of Ambegaokar and Halperin\cite{ambe69} (AH) for a resistively-shunted Josephson junction\cite{mccu68} plus thermally-driven current fluctuations.  The model exhibits linear resistance at small currents and a rapid rise towards the normal state resistance as the Josephson critical current is approached.

\begin{figure}[b]
\centerline{\includegraphics[width=3.0in]{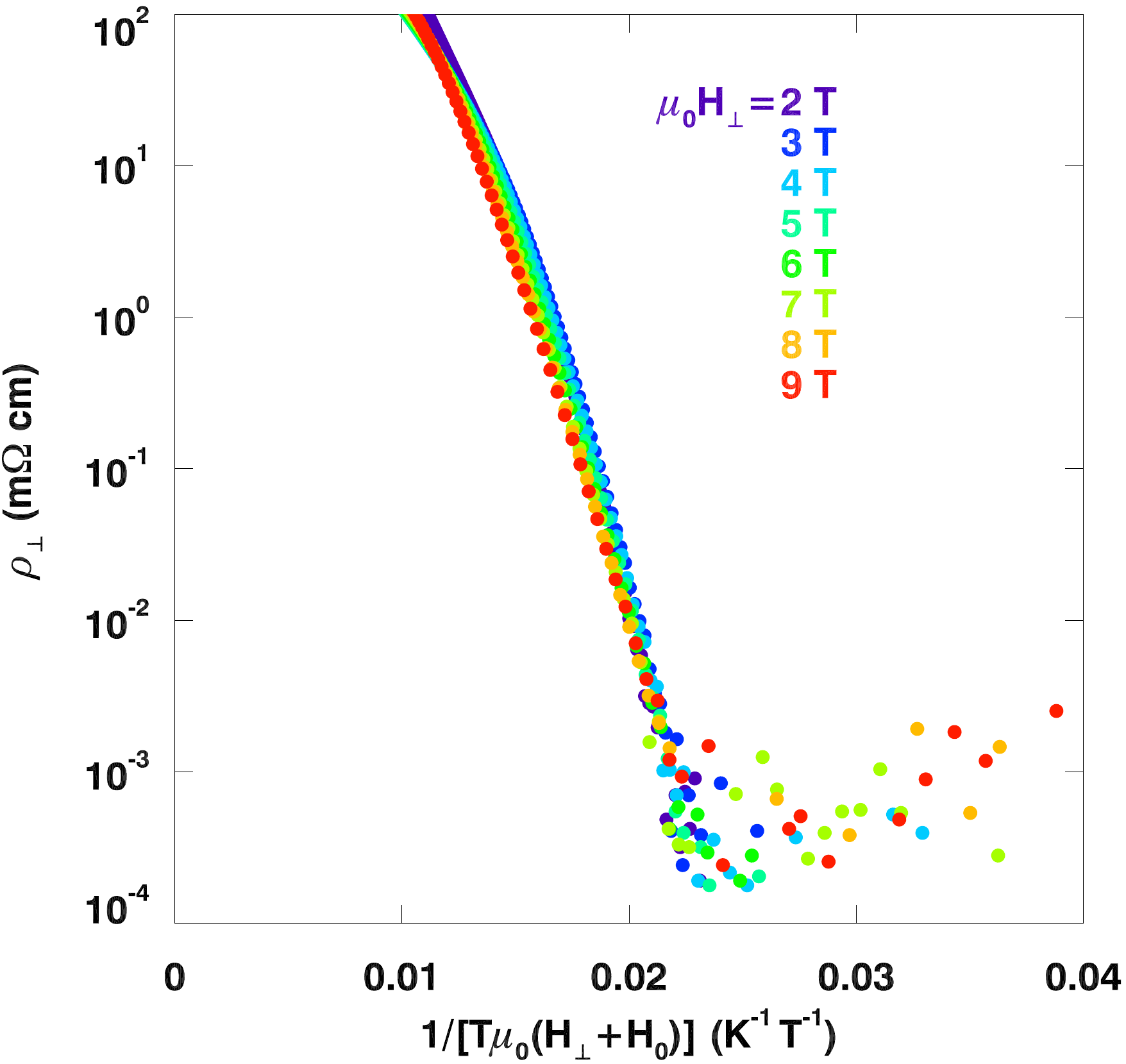}}
\caption{ (Color online) Data for $\rho_\bot(T)$ with $\mu_0H_\bot = 2$ to 9~T plotted vs.\ $1/[T\mu_0(H_\bot+H_0)]$, with $\mu_0H_0=2.2$~T.
\label{fg:scale}}
\end{figure}

A number of previous studies of cuprates have invoked the AH results,\cite{ambe69} proposing that the temperature and field dependence of $\rho_\bot$ can be described by treating each crystal as a stack of independent interlayer Josephson junctions.\cite{bric91,gray93,hett95,suzu98b}  One issue is that the sensitivity to thermal noise depends on the extensive critical current, and hence depends on the effective junction area, which in practice can be much smaller than the sample cross section.  Hettinger {\it et al.}\cite{hett95} demonstrated experimentally that the effective area is given by $\phi_0/(B_\bot+B_{\bot 0})$, where $\phi_0$ is the magnetic flux quantum and $B_0$ is a parameter that characterizes the behavior at zero field.  To test this in our case, we note that the AH formula for resistivity measured with a current less than the critical value is
\begin{equation}
 \rho_\bot   = \rho^*_\bot/ [I_0(\gamma_0/2)]^{2},
 \label{eq:rhobot}
\end{equation}
where $\rho^*_\bot$ is an effective normal state resistivity (with corrections due to quasiparticle flow in parallel with Cooper pairs\cite{hett95,laty99}), $I_0$ is the same modified Bessel function as in Eq.~(1), and
\begin{equation}
 \gamma_0 = E_{\rm J} / k_{\rm B}T.
\end{equation}
According to Hettinger {\it et al.},\cite{hett95} the Josephson-coupling energy $E_{\rm J}$ can be written as
\begin{equation}
 E_{\rm J} = e_{\rm J} \phi_0 / (B_\bot+B_{\bot 0}),
 \label{eq:EJ}
\end{equation}
where $e_{\rm J}$ is the Josephson energy density, proportional to the critical current density.
Assuming $B_\bot\approx\mu_0H_\bot$ and ignoring temperature variations of $\rho^*_\bot$ at low temperatures, $\rho_\bot$ should scale as $[T\mu_0(H_\bot+H_{\bot0})]^{-1}$.  We show in Fig.~\ref{fg:scale} that this scaling works rather well for 2~T~$\le \mu_0H_\bot\le9$~T if we take $\mu_0H_{\bot0}=2.2$~T.   This analysis involves measurements in the temperature range of 5 to 15~K, where optical measurements\cite{home12} of the Josephson coupling suggest that there should be relatively little temperature dependence of $E_{\rm J}$.  The scaling becomes poorer if we include that data for $\mu_0H_\bot=1$~T.

To evaluate the magnitude of $E_{\rm J}$, we fit Eq.~(\ref{eq:rhobot}) to the data for $\rho_\bot(9\,\mbox{T}) < 1\times 10^{-2}$ m$\Omega$ cm, obtaining $E_{\rm J}(9\,\mbox{T}) = 8.2\pm0.5$~meV, corresponding to the field-independent quantity $e_{\rm J}=44$~eV/$\mu$m$^2$.   The Josephson coupling can also be expressed in terms of the magnetic penetration depth $\lambda_\bot$, which, according to the analysis of optical reflectivity measurements,\cite{home12} has a low temperature value of 13.4~$\mu$m.  Using the formula\cite{clem91,hett95}
\begin{equation}
 \lambda_\bot^2 = \phi_0^2/8\pi^3se_{\rm J},
\end{equation}
where $s=6.6$~\AA\ is the layer separation, we find that our value of $e_{\rm J}$ corresponds to $\lambda_\bot=6.1~\mu$m, about a factor of two smaller than the optical value.  We consider this to be rather good agreement, given the possibility that there could be a scale factor for the effective area identified by Hettinger {\it et al.}\cite{hett95}   Furthermore, our field dependence of $E_{\rm J}$ is similar to that determined by Schafgans {\it et al.}\cite{scha10} in LSCO $x=0.10$ (where $\lambda_\bot=12.6~\mu$m).

\subsection{Discussion}

We have presented evidence that, for our LBCO $x=0.095$ sample, there is linear resistivity perpendicular to the layers under conditions where there is no linear in-plane resistivity.  These conditions include substantial magnetic fields perpendicular to the layers.  Linear $\rho_\bot$ can appear for measuring currents well below the effective Josephson critical current.  The latter behavior is consistent with previous studies\cite{bric91,gray93,hett95,suzu98b} in which $\rho_\bot$ was analyzed in terms of a stack of independent Josephson junctions, with resistivity arising from thermal fluctuations.\cite{ambe69}  Our demonstration of finite $\rho_\bot$ with no linear $\rho_\|$ goes beyond that earlier work.

When $\rho_\bot$ is finite, it indicates a lack of superconducting phase coherence perpendicular to the layers.  In the mixed phase, the loss of phase coherence between the layers is expected to correspond to loss of coherence within the layers.\cite{kosh96a}  This expectation is associated with the idea that the loss of coherence should be tied to fluctuations of pancake vortices.  Interlayer interactions may help to pin the vortices, but once interlayer coherence is lost, the pancake vortices are expected to become unpinned, resulting in the loss of superconducting order.  Our experimental results suggest that some sort of vortex glass state survives within the layers, despite the loss of interlayer coherence.

A similar situation to the one found here was previously observed in LBCO $x=1/8$ in zero field.\cite{li07,tran08}  In the latter case, superconducting order was detected parallel to the planes while $\rho_\bot$ was still substantial.  That phase was also found to survive in finite $H_\bot$.  The unusual behavior in the $x=1/8$ sample is closely associated with the occurrence of spin and charge stripe order.  Thus, it is relevant to probe the impact of $H_\bot$ on stripe correlations in the present sample.

\section{Diffraction measurements of stripe order}
\label{sc:diff}

We have previously used neutron and x-ray diffraction to characterize the spin and charge ordering transitions, as well as the structural transitions in LBCO.\cite{huck11}  In zero-field, the spin and charge order diffraction peak intensities for the $x=0.095$ composition are reduced by an order of magnitude compared to those in the $x=0.125$ composition, where stripe order is maximized.  Here, our main focus is on the impact of $H_\bot$.

\begin{figure}[t]
\centerline{\includegraphics[width=3.0in]{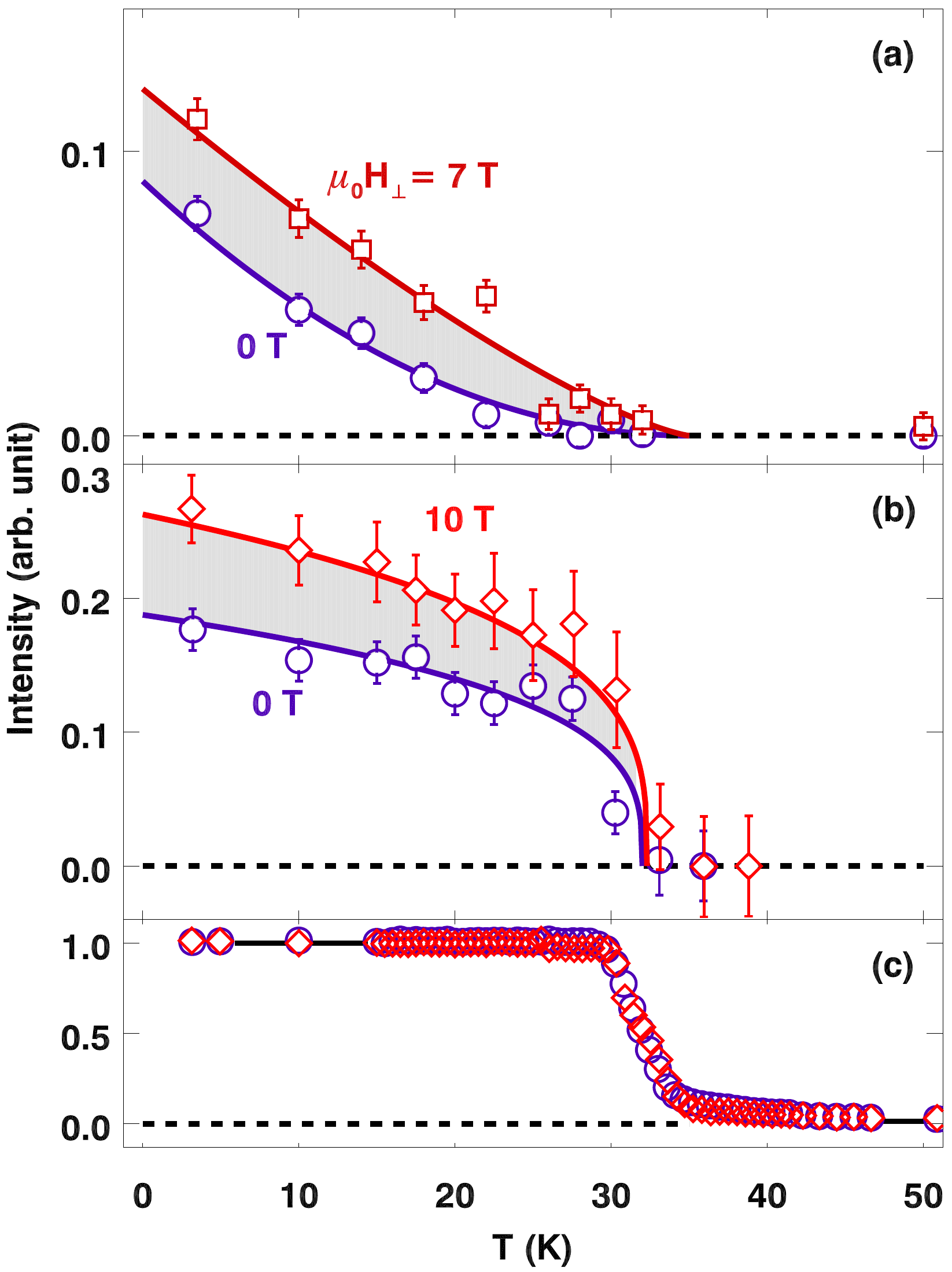}}
\caption{ (Color online)
(a) Integrated intensity of the magnetic superlattice peak at wave vector $(0.6,0.5,0)$ in $\mu_0 H_\bot = 0$~T (violet circles) and 7~T (red squares), obtained by neutron diffraction.
(b)  Integrated intensity of the charge-order superlattice peak $(0.2,0,8.5)$ in $\mu_0 H_\bot = 0$~T (violet circles) and 10~T (red diamonds), obtained by x-ray diffraction.
(c) Integrated intensity of the (300) superlattice peak, characterizing the structural transition to the low-temperature phase, in 0~T and 10~T as in (b).  For (a) and (b), intensities are normalized (approximately) to results for LBCO $x=0.125$ in zero field\cite{huck10}; for (c), intensities are normalized at low temperature.  Error bars correspond to one standard deviation of counting statistics.  Lines through the data points are guides to the eye.  Gray regions emphasize the change induced by the magnetic field.
\label{fg:diff}  }
\end{figure}

For reference, the temperature dependence of a structural superlattice peak characteristic of the LTLO phase, measured by x-ray diffraction, is illustrated in Fig.~\ref{fg:diff}(c).  Correlated with the appearance of this phase are superlattice peaks associated with spin and charge stripe order, as shown in Fig.~\ref{fg:diff}(a) and (b), respectively.  Application of a significant $H_\bot$ results in a substantial growth of the stripe order intensity.  While boosting the spin order by an applied magnetic field is not new,\cite{lake02,chan08,fuji12} this is the first observation, to our knowledge, of field-enhanced charge-stripe order.  

The stripe order develops only at temperatures below the zero-field superconducting $T_c$.  In this regime, the substantial $H_\bot$ penetrates the CuO$_2$ layers as flux quanta.  Hence, the induced stripe order is likely associated with the superconducting vortex cores,\cite{lake01,lake02} consistent with the implications of scanning tunneling spectroscopic observations.\cite{hoff02}  The stripe correlation length\cite{huck11,wen12a} of $\sim100$~\AA\ is significantly larger than the typical vortex core size, indicating that the static charge and spin stripes coexist with the superconductivity in halo regions.\cite{kive02}  The LTLO structure helps to stabilize the stripe order in our sample, although the observations in LSCO demonstrate that stripe order can also be induced in the LTO phase.\cite{lake02,chan08}

The LTO-LTLO/LTT transition is rather sensitive to composition.  The crystal used for the x-ray measurements has its transition completed on cooling to 30~K, as indicated in Fig.~\ref{fg:diff}, whereas the crystals studied\cite{wen12a} by neutron diffraction and transport measurements have the transition end at 27~K.  Based on the phase diagram presented in Ref.~\onlinecite{huck11}, this difference of 3~K corresponds to a composition difference of $\sim0.002$.  This sensitivity is also relevant to comparisons with work by other groups.  Dunsiger {\it et al.}\cite{duns08} studied a sample with the same nominal composition; however, the LTO-LTT transition was at 45~K, consistent with a Ba composition greater than ours by $\sim0.015$.  The fact that they did not see any change in the intensity of a spin order superlattice peak in $\mu_0H_\bot=7$~T is consistent with the greater hole concentration and larger zero-field stripe order.  A significant field enhancement is only observed when the zero-field intensity is weak.\cite{chan08,fuji12}

\section{Summary and Discussion}
\label{sc:sum}

We have experimentally studied the anisotropic resistivity in LBCO $x=0.095$ and the impact of an applied magnetic field, especially when oriented perpendicular to the CuO$_2$ layers.  Measuring the onset of detectable resistivity vs.\ temperature for various values of $H_\bot$, we find quite different thresholds for the loss of superconducting order depending on whether the measurement current is parallel or perpendicular to the layers.  For current parallel to the layers, the nonlinear voltage vs.\ current behavior is consistent with survival of superconducting order to rather high temperatures and magnetic fields.  In contrast, a small current applied perpendicular to the layers leads to linear voltage dependence (within the regime of detectable resistivity), with nonlinear behavior at higher currents.  This behavior is consistent with thermal-noise-induced voltage fluctuations across independent interlayer Josephson junctions.\cite{ambe69,bric91,gray93,hett95,suzu98b}  This effect is expected to be quite sensitive to the effective area of the Josephson junction.\cite{ambe69}  Applying the empirical result of Hettinger {\it et al.}\cite{hett95} that the effective area is comparable to the area per vortex, we get reasonable consistency with the optical measurement of $\lambda_\bot$.\cite{home12}

We have also used diffraction techniques to measure the impact of $H_\bot$ on stripe order.  We have observed that both charge and spin stripe orders are significantly enhanced by the field.    The correlation lengths for the vortex-induced stripe orders\cite{wen12a} are significantly larger than the typical vortex core size, so that the induced stripe order must coexist with superconducting screening currents.

Can we find a way to make sense of the different behaviors in $\rho_{\|}$ and $\rho_{\bot}$ in the regime where one indicates superconducting order and the other does not?  One way to possibly understand these differences is in terms of anisotropic vortex pinning.  In measuring $\rho_{\|}$, the response necessarily indicates that vortices are pinned.  This pinning may be aided by the stripes present in the system.  In the LTLO or LTT lattice structures, stripes are pinned in orthogonal directions from one layer to the next.  For a current flowing parallel to the layers with applied field $H_\bot$, the transverse Lorentz force in the plane would push pancake vortices in half of the layers in the modulation direction of the induced stripe order.  Given that the vortices appear to induce stripe order, the Lorentz force may serve to pin each vortex core within the halo of induced spin order.  Because vortices between layers are (attractively) coupled electromagnetically,\cite{rama09} pinning in one layer will aid pinning in adjacent layers, even in those layers where the Lorentz force acts along the stripe.  In contrast, for a current perpendicular to the layers, there is no Lorentz force on the pancake vortices, thus allowing them freedom to fluctuate parallel to the stripes, resulting in a lower threshold for detectable resistivity.

Another possibility to consider is whether there might be an intimate connection between the superconductivity and stripe order.  A state with apparent superconducting order parallel to the planes but finite resistivity between the planes was previously observed in LBCO $x=1/8$.\cite{li07,tran08}  There it occurs in zero field and onsets together with spin stripe order.  Frustration of the interlayer Josephson coupling is evident from the extreme anisotropy of the resistivity and diamagnetism.\cite{tran08}  For the present case, $x=0.095$, the interlayer Josephson coupling is finite in zero field, but the coupling is reduced by the field.  Dissipation appears while the Josephson coupling is still finite.  A large magnitude of $\rho_\bot$ in strong $H_\bot$ occurs for a range of temperatures where $\rho_\|$ remains very small, suggesting a frustration of Josephson coherence.  We have also observed that $H_\bot$ enhances spin and charge stripe order.  Thus, there are significant qualitatively similarities between the phases of superconductivity with uniaxial resistivity in the $x=0.095$ and 0.125 samples.  Of course, there are also some quantitative differences.  Even in the state of uniaxial resistivity, the in-plane superconductivity appears much more robust to $H_\bot$ in the $x=0.095$ sample than for $x=1/8$.  

The pair-density-wave (PDW) superconducting state has been proposed\cite{hime02,berg07,berg09b} to explain the frustrated Josephson coupling in cuprates with strong stripe order.\cite{li07,taji01,ichi00}  In the PDW state, the pair wave function is intertwined with the spin and charge stripe order such that the spin order and pair wave function minimize their overlap.  While there has not been definitive evidence for the PDW state, there have been recent observations of time-reversal-symmetry breaking associated with the onset of the charge stripe order,\cite{li11,kara12} which provide possible connections.  In any case, one expects this state to be sensitive to disorder and perturbations, and the more rapid suppression of the superconducting signatures in $H_\bot$ in LBCO $x=1/8$ is qualitatively consistent with expectations. 

The coexistence of uniform superconducting order with the PDW is expected to eliminate the sensitivity to disorder,\cite{berg09b} and this might help to explain the more robust field dependence of the in-plane superconductivity in LBCO $x=0.095$.  Of course, the presence of uniform superconducting order should also provide a channel for interlayer Josephson coupling, which is certainly a significant factor for the $x=0.095$ sample.  Nevertheless, even if the PDW state is relevant,  we are not aware of any theory that could explain a superconducting state with field-induced uniaxial resistivity.  The experimental parallels between the unusual phase found in both the $x=0.095$ and 0.125 samples should provide some guidance to attempts to understand the phase theoretically.

\acknowledgments

We gratefully acknowledge discussions with S. A. Kivelson, E. Fradkin, A. E. Koshelev, V. Oganesyan, A. Tsvelik, and G. Y. Xu.  JSW, QJ, SJH, and ZJX were supported by the Center for Emergent Superconductivity, an Energy Frontier Research Center funded by the Office of Basic Energy Sciences (BES), Division of Materials Science and Engineering, U.S. Department of Energy.  Other work at Brookhaven is supported by BES through Contract No.\ DE-AC02-98CH10886.   SPINS at NCNR is supported by the National Science Foundation under Agreement No.\ DMR-0454672.


\begin{thebibliography}{59}%
\makeatletter
\providecommand \@ifxundefined [1]{%
 \@ifx{#1\undefined}
}%
\providecommand \@ifnum [1]{%
 \ifnum #1\expandafter \@firstoftwo
 \else \expandafter \@secondoftwo
 \fi
}%
\providecommand \@ifx [1]{%
 \ifx #1\expandafter \@firstoftwo
 \else \expandafter \@secondoftwo
 \fi
}%
\providecommand \natexlab [1]{#1}%
\providecommand \enquote  [1]{``#1''}%
\providecommand \bibnamefont  [1]{#1}%
\providecommand \bibfnamefont [1]{#1}%
\providecommand \citenamefont [1]{#1}%
\providecommand \href@noop [0]{\@secondoftwo}%
\providecommand \href [0]{\begingroup \@sanitize@url \@href}%
\providecommand \@href[1]{\@@startlink{#1}\@@href}%
\providecommand \@@href[1]{\endgroup#1\@@endlink}%
\providecommand \@sanitize@url [0]{\catcode `\\12\catcode `\$12\catcode
  `\&12\catcode `\#12\catcode `\^12\catcode `\_12\catcode `\%12\relax}%
\providecommand \@@startlink[1]{}%
\providecommand \@@endlink[0]{}%
\providecommand \url  [0]{\begingroup\@sanitize@url \@url }%
\providecommand \@url [1]{\endgroup\@href {#1}{\urlprefix }}%
\providecommand \urlprefix  [0]{URL }%
\providecommand \Eprint [0]{\href }%
\providecommand \doibase [0]{http://dx.doi.org/}%
\providecommand \selectlanguage [0]{\@gobble}%
\providecommand \bibinfo  [0]{\@secondoftwo}%
\providecommand \bibfield  [0]{\@secondoftwo}%
\providecommand \translation [1]{[#1]}%
\providecommand \BibitemOpen [0]{}%
\providecommand \bibitemStop [0]{}%
\providecommand \bibitemNoStop [0]{.\EOS\space}%
\providecommand \EOS [0]{\spacefactor3000\relax}%
\providecommand \BibitemShut  [1]{\csname bibitem#1\endcsname}%
\let\auto@bib@innerbib\@empty
\bibitem [{\citenamefont {Taillefer}(2009)}]{tail09}%
  \BibitemOpen
  \bibfield  {author} {\bibinfo {author} {\bibfnamefont {L.}~\bibnamefont
  {Taillefer}},\ }\href@noop {} {\bibfield  {journal} {\bibinfo  {journal} {J.
  Phys.: Condens. Matter}\ }\textbf {\bibinfo {volume} {21}},\ \bibinfo {pages}
  {164212} (\bibinfo {year} {2009})}\BibitemShut {NoStop}%
\bibitem [{\citenamefont {Sebastian}\ \emph {et~al.}(2011)\citenamefont
  {Sebastian}, \citenamefont {Lonzarich},\ and\ \citenamefont
  {Harrison}}]{seba12}%
  \BibitemOpen
  \bibfield  {author} {\bibinfo {author} {\bibfnamefont {S.~E.}\ \bibnamefont
  {Sebastian}}, \bibinfo {author} {\bibfnamefont {G.~G.}\ \bibnamefont
  {Lonzarich}}, \ and\ \bibinfo {author} {\bibfnamefont {N.}~\bibnamefont
  {Harrison}},\ }\href@noop {} {\enquote {\bibinfo {title} {{Towards resolution
  of the Fermi surface in underdoped high-$T_c$ superconductors}},}\ }
  (\bibinfo {year} {2011}),\ \Eprint {http://arxiv.org/abs/arXiv:1112.1373}
  {arXiv:1112.1373} \BibitemShut {NoStop}%
\bibitem [{\citenamefont {Kleiner}\ \emph {et~al.}(1992)\citenamefont
  {Kleiner}, \citenamefont {Steinmeyer}, \citenamefont {Kunkel},\ and\
  \citenamefont {M\"uller}}]{klei92}%
  \BibitemOpen
  \bibfield  {author} {\bibinfo {author} {\bibfnamefont {R.}~\bibnamefont
  {Kleiner}}, \bibinfo {author} {\bibfnamefont {F.}~\bibnamefont {Steinmeyer}},
  \bibinfo {author} {\bibfnamefont {G.}~\bibnamefont {Kunkel}}, \ and\ \bibinfo
  {author} {\bibfnamefont {P.}~\bibnamefont {M\"uller}},\ }\href@noop {}
  {\bibfield  {journal} {\bibinfo  {journal} {Phys. Rev. Lett.}\ }\textbf
  {\bibinfo {volume} {68}},\ \bibinfo {pages} {2394} (\bibinfo {year}
  {1992})}\BibitemShut {NoStop}%
\bibitem [{\citenamefont {Uematsu}\ \emph {et~al.}(1998)\citenamefont
  {Uematsu}, \citenamefont {Nakajima}, \citenamefont {Yamashita}, \citenamefont
  {Tanaka},\ and\ \citenamefont {Kojima}}]{uema98}%
  \BibitemOpen
  \bibfield  {author} {\bibinfo {author} {\bibfnamefont {Y.}~\bibnamefont
  {Uematsu}}, \bibinfo {author} {\bibfnamefont {K.}~\bibnamefont {Nakajima}},
  \bibinfo {author} {\bibfnamefont {T.}~\bibnamefont {Yamashita}}, \bibinfo
  {author} {\bibfnamefont {I.}~\bibnamefont {Tanaka}}, \ and\ \bibinfo {author}
  {\bibfnamefont {H.}~\bibnamefont {Kojima}},\ }\href {\doibase
  10.1063/1.122601} {\bibfield  {journal} {\bibinfo  {journal} {Applied Physics
  Letters}\ }\textbf {\bibinfo {volume} {73}},\ \bibinfo {pages} {2820}
  (\bibinfo {year} {1998})}\BibitemShut {NoStop}%
\bibitem [{\citenamefont {Clem}(1991)}]{clem91}%
  \BibitemOpen
  \bibfield  {author} {\bibinfo {author} {\bibfnamefont {J.~R.}\ \bibnamefont
  {Clem}},\ }\href@noop {} {\bibfield  {journal} {\bibinfo  {journal} {Phys.
  Rev. B}\ }\textbf {\bibinfo {volume} {43}},\ \bibinfo {pages} {7837}
  (\bibinfo {year} {1991})}\BibitemShut {NoStop}%
\bibitem [{\citenamefont {Glazman}\ and\ \citenamefont
  {Koshelev}(1991)}]{glaz91}%
  \BibitemOpen
  \bibfield  {author} {\bibinfo {author} {\bibfnamefont {L.~I.}\ \bibnamefont
  {Glazman}}\ and\ \bibinfo {author} {\bibfnamefont {A.~E.}\ \bibnamefont
  {Koshelev}},\ }\href@noop {} {\bibfield  {journal} {\bibinfo  {journal}
  {Phys. Rev. B}\ }\textbf {\bibinfo {volume} {43}},\ \bibinfo {pages} {2835}
  (\bibinfo {year} {1991})}\BibitemShut {NoStop}%
\bibitem [{\citenamefont {Daemen}\ \emph {et~al.}(1993)\citenamefont {Daemen},
  \citenamefont {Bulaevskii}, \citenamefont {Maley},\ and\ \citenamefont
  {Coulter}}]{daem93}%
  \BibitemOpen
  \bibfield  {author} {\bibinfo {author} {\bibfnamefont {L.~L.}\ \bibnamefont
  {Daemen}}, \bibinfo {author} {\bibfnamefont {L.~N.}\ \bibnamefont
  {Bulaevskii}}, \bibinfo {author} {\bibfnamefont {M.~P.}\ \bibnamefont
  {Maley}}, \ and\ \bibinfo {author} {\bibfnamefont {J.~Y.}\ \bibnamefont
  {Coulter}},\ }\href@noop {} {\bibfield  {journal} {\bibinfo  {journal} {Phys.
  Rev. Lett.}\ }\textbf {\bibinfo {volume} {70}},\ \bibinfo {pages} {1167}
  (\bibinfo {year} {1993})}\BibitemShut {NoStop}%
\bibitem [{\citenamefont {Bulaevskii}\ \emph {et~al.}(1995)\citenamefont
  {Bulaevskii}, \citenamefont {Maley},\ and\ \citenamefont {Tachiki}}]{bula95}%
  \BibitemOpen
  \bibfield  {author} {\bibinfo {author} {\bibfnamefont {L.~N.}\ \bibnamefont
  {Bulaevskii}}, \bibinfo {author} {\bibfnamefont {M.~P.}\ \bibnamefont
  {Maley}}, \ and\ \bibinfo {author} {\bibfnamefont {M.}~\bibnamefont
  {Tachiki}},\ }\href@noop {} {\bibfield  {journal} {\bibinfo  {journal} {Phys.
  Rev. Lett.}\ }\textbf {\bibinfo {volume} {74}},\ \bibinfo {pages} {801}
  (\bibinfo {year} {1995})}\BibitemShut {NoStop}%
\bibitem [{\citenamefont {Koshelev}\ \emph {et~al.}(1996)\citenamefont
  {Koshelev}, \citenamefont {Glazman},\ and\ \citenamefont {Larkin}}]{kosh96b}%
  \BibitemOpen
  \bibfield  {author} {\bibinfo {author} {\bibfnamefont {A.~E.}\ \bibnamefont
  {Koshelev}}, \bibinfo {author} {\bibfnamefont {L.~I.}\ \bibnamefont
  {Glazman}}, \ and\ \bibinfo {author} {\bibfnamefont {A.~I.}\ \bibnamefont
  {Larkin}},\ }\href@noop {} {\bibfield  {journal} {\bibinfo  {journal} {Phys.
  Rev. B}\ }\textbf {\bibinfo {volume} {53}},\ \bibinfo {pages} {2786}
  (\bibinfo {year} {1996})}\BibitemShut {NoStop}%
\bibitem [{\citenamefont {Fisher}\ \emph {et~al.}(1991)\citenamefont {Fisher},
  \citenamefont {Fisher},\ and\ \citenamefont {Huse}}]{fish91}%
  \BibitemOpen
  \bibfield  {author} {\bibinfo {author} {\bibfnamefont {D.~S.}\ \bibnamefont
  {Fisher}}, \bibinfo {author} {\bibfnamefont {M.~P.~A.}\ \bibnamefont
  {Fisher}}, \ and\ \bibinfo {author} {\bibfnamefont {D.~A.}\ \bibnamefont
  {Huse}},\ }\href@noop {} {\bibfield  {journal} {\bibinfo  {journal} {Phys.
  Rev. B}\ }\textbf {\bibinfo {volume} {43}},\ \bibinfo {pages} {130} (\bibinfo
  {year} {1991})}\BibitemShut {NoStop}%
\bibitem [{\citenamefont {Koshelev}(1996)}]{kosh96a}%
  \BibitemOpen
  \bibfield  {author} {\bibinfo {author} {\bibfnamefont {A.~E.}\ \bibnamefont
  {Koshelev}},\ }\href@noop {} {\bibfield  {journal} {\bibinfo  {journal}
  {Phys. Rev. Lett.}\ }\textbf {\bibinfo {volume} {76}},\ \bibinfo {pages}
  {1340} (\bibinfo {year} {1996})}\BibitemShut {NoStop}%
\bibitem [{\citenamefont {Oganesyan}\ \emph {et~al.}(2006)\citenamefont
  {Oganesyan}, \citenamefont {Huse},\ and\ \citenamefont {Sondhi}}]{ogan06}%
  \BibitemOpen
  \bibfield  {author} {\bibinfo {author} {\bibfnamefont {V.}~\bibnamefont
  {Oganesyan}}, \bibinfo {author} {\bibfnamefont {D.~A.}\ \bibnamefont {Huse}},
  \ and\ \bibinfo {author} {\bibfnamefont {S.~L.}\ \bibnamefont {Sondhi}},\
  }\href@noop {} {\bibfield  {journal} {\bibinfo  {journal} {Phys. Rev. B}\
  }\textbf {\bibinfo {volume} {73}},\ \bibinfo {pages} {094503} (\bibinfo
  {year} {2006})}\BibitemShut {NoStop}%
\bibitem [{\citenamefont {Duli\'c}\ \emph {et~al.}(2001)\citenamefont
  {Duli\'c}, \citenamefont {Hak}, \citenamefont {van~der Marel}, \citenamefont
  {Hardy}, \citenamefont {Koshelev}, \citenamefont {Liang}, \citenamefont
  {Bonn},\ and\ \citenamefont {Willemsen}}]{duli01}%
  \BibitemOpen
  \bibfield  {author} {\bibinfo {author} {\bibfnamefont {D.}~\bibnamefont
  {Duli\'c}}, \bibinfo {author} {\bibfnamefont {S.~J.}\ \bibnamefont {Hak}},
  \bibinfo {author} {\bibfnamefont {D.}~\bibnamefont {van~der Marel}}, \bibinfo
  {author} {\bibfnamefont {W.~N.}\ \bibnamefont {Hardy}}, \bibinfo {author}
  {\bibfnamefont {A.~E.}\ \bibnamefont {Koshelev}}, \bibinfo {author}
  {\bibfnamefont {R.}~\bibnamefont {Liang}}, \bibinfo {author} {\bibfnamefont
  {D.~A.}\ \bibnamefont {Bonn}}, \ and\ \bibinfo {author} {\bibfnamefont
  {B.~A.}\ \bibnamefont {Willemsen}},\ }\href {\doibase
  10.1103/PhysRevLett.86.4660} {\bibfield  {journal} {\bibinfo  {journal}
  {Phys. Rev. Lett.}\ }\textbf {\bibinfo {volume} {86}},\ \bibinfo {pages}
  {4660} (\bibinfo {year} {2001})}\BibitemShut {NoStop}%
\bibitem [{\citenamefont {Schafgans}\ \emph {et~al.}(2010)\citenamefont
  {Schafgans}, \citenamefont {LaForge}, \citenamefont {Dordevic}, \citenamefont
  {Qazilbash}, \citenamefont {Padilla}, \citenamefont {Burch}, \citenamefont
  {Li}, \citenamefont {Komiya}, \citenamefont {Ando},\ and\ \citenamefont
  {Basov}}]{scha10}%
  \BibitemOpen
  \bibfield  {author} {\bibinfo {author} {\bibfnamefont {A.~A.}\ \bibnamefont
  {Schafgans}}, \bibinfo {author} {\bibfnamefont {A.~D.}\ \bibnamefont
  {LaForge}}, \bibinfo {author} {\bibfnamefont {S.~V.}\ \bibnamefont
  {Dordevic}}, \bibinfo {author} {\bibfnamefont {M.~M.}\ \bibnamefont
  {Qazilbash}}, \bibinfo {author} {\bibfnamefont {W.~J.}\ \bibnamefont
  {Padilla}}, \bibinfo {author} {\bibfnamefont {K.~S.}\ \bibnamefont {Burch}},
  \bibinfo {author} {\bibfnamefont {Z.~Q.}\ \bibnamefont {Li}}, \bibinfo
  {author} {\bibfnamefont {S.}~\bibnamefont {Komiya}}, \bibinfo {author}
  {\bibfnamefont {Y.}~\bibnamefont {Ando}}, \ and\ \bibinfo {author}
  {\bibfnamefont {D.~N.}\ \bibnamefont {Basov}},\ }\href@noop {} {\bibfield
  {journal} {\bibinfo  {journal} {Phys. Rev. Lett.}\ }\textbf {\bibinfo
  {volume} {104}},\ \bibinfo {pages} {157002} (\bibinfo {year}
  {2010})}\BibitemShut {NoStop}%
\bibitem [{\citenamefont {Lake}\ \emph {et~al.}(2002)\citenamefont {Lake},
  \citenamefont {{R\o nnow}}, \citenamefont {Christensen}, \citenamefont
  {Aeppli}, \citenamefont {Lefmann}, \citenamefont {McMorrow}, \citenamefont
  {Vorderwisch}, \citenamefont {Smeibidl}, \citenamefont {Mangkorntong},
  \citenamefont {Sasagawa}, \citenamefont {Nohara}, \citenamefont {Takagi},\
  and\ \citenamefont {Mason}}]{lake02}%
  \BibitemOpen
  \bibfield  {author} {\bibinfo {author} {\bibfnamefont {B.}~\bibnamefont
  {Lake}}, \bibinfo {author} {\bibfnamefont {H.~M.}\ \bibnamefont {{R\o
  nnow}}}, \bibinfo {author} {\bibfnamefont {N.~B.}\ \bibnamefont
  {Christensen}}, \bibinfo {author} {\bibfnamefont {G.}~\bibnamefont {Aeppli}},
  \bibinfo {author} {\bibfnamefont {K.}~\bibnamefont {Lefmann}}, \bibinfo
  {author} {\bibfnamefont {D.~F.}\ \bibnamefont {McMorrow}}, \bibinfo {author}
  {\bibfnamefont {P.}~\bibnamefont {Vorderwisch}}, \bibinfo {author}
  {\bibfnamefont {P.}~\bibnamefont {Smeibidl}}, \bibinfo {author}
  {\bibfnamefont {N.}~\bibnamefont {Mangkorntong}}, \bibinfo {author}
  {\bibfnamefont {T.}~\bibnamefont {Sasagawa}}, \bibinfo {author}
  {\bibfnamefont {M.}~\bibnamefont {Nohara}}, \bibinfo {author} {\bibfnamefont
  {H.}~\bibnamefont {Takagi}}, \ and\ \bibinfo {author} {\bibfnamefont {T.~E.}\
  \bibnamefont {Mason}},\ }\href@noop {} {\bibfield  {journal} {\bibinfo
  {journal} {Nature}\ }\textbf {\bibinfo {volume} {415}},\ \bibinfo {pages}
  {299} (\bibinfo {year} {2002})}\BibitemShut {NoStop}%
\bibitem [{\citenamefont {Chang}\ \emph {et~al.}(2008)\citenamefont {Chang},
  \citenamefont {Niedermayer}, \citenamefont {Gilardi}, \citenamefont
  {Christensen}, \citenamefont {Ronnow}, \citenamefont {McMorrow},
  \citenamefont {Ay}, \citenamefont {Stahn}, \citenamefont {Sobolev},
  \citenamefont {Hiess}, \citenamefont {Pailhes}, \citenamefont {Baines},
  \citenamefont {Momono}, \citenamefont {Oda}, \citenamefont {Ido},\ and\
  \citenamefont {Mesot}}]{chan08}%
  \BibitemOpen
  \bibfield  {author} {\bibinfo {author} {\bibfnamefont {J.}~\bibnamefont
  {Chang}}, \bibinfo {author} {\bibfnamefont {C.}~\bibnamefont {Niedermayer}},
  \bibinfo {author} {\bibfnamefont {R.}~\bibnamefont {Gilardi}}, \bibinfo
  {author} {\bibfnamefont {N.}~\bibnamefont {Christensen}}, \bibinfo {author}
  {\bibfnamefont {H.}~\bibnamefont {Ronnow}}, \bibinfo {author} {\bibfnamefont
  {D.}~\bibnamefont {McMorrow}}, \bibinfo {author} {\bibfnamefont
  {M.}~\bibnamefont {Ay}}, \bibinfo {author} {\bibfnamefont {J.}~\bibnamefont
  {Stahn}}, \bibinfo {author} {\bibfnamefont {O.}~\bibnamefont {Sobolev}},
  \bibinfo {author} {\bibfnamefont {A.}~\bibnamefont {Hiess}}, \bibinfo
  {author} {\bibfnamefont {S.}~\bibnamefont {Pailhes}}, \bibinfo {author}
  {\bibfnamefont {C.}~\bibnamefont {Baines}}, \bibinfo {author} {\bibfnamefont
  {N.}~\bibnamefont {Momono}}, \bibinfo {author} {\bibfnamefont
  {M.}~\bibnamefont {Oda}}, \bibinfo {author} {\bibfnamefont {M.}~\bibnamefont
  {Ido}}, \ and\ \bibinfo {author} {\bibfnamefont {J.}~\bibnamefont {Mesot}},\
  }\href@noop {} {\bibfield  {journal} {\bibinfo  {journal} {Phys. Rev. B}\
  }\textbf {\bibinfo {volume} {78}},\ \bibinfo {pages} {104525} (\bibinfo
  {year} {2008})}\BibitemShut {NoStop}%
\bibitem [{\citenamefont {Fujita}\ \emph {et~al.}(2012)\citenamefont {Fujita},
  \citenamefont {Hiraka}, \citenamefont {Matsuda}, \citenamefont {Matsuura},
  \citenamefont {Tranquada}, \citenamefont {Wakimoto}, \citenamefont {Xu},\
  and\ \citenamefont {Yamada}}]{fuji12}%
  \BibitemOpen
  \bibfield  {author} {\bibinfo {author} {\bibfnamefont {M.}~\bibnamefont
  {Fujita}}, \bibinfo {author} {\bibfnamefont {H.}~\bibnamefont {Hiraka}},
  \bibinfo {author} {\bibfnamefont {M.}~\bibnamefont {Matsuda}}, \bibinfo
  {author} {\bibfnamefont {M.}~\bibnamefont {Matsuura}}, \bibinfo {author}
  {\bibfnamefont {J.~M.}\ \bibnamefont {Tranquada}}, \bibinfo {author}
  {\bibfnamefont {S.}~\bibnamefont {Wakimoto}}, \bibinfo {author}
  {\bibfnamefont {G.}~\bibnamefont {Xu}}, \ and\ \bibinfo {author}
  {\bibfnamefont {K.}~\bibnamefont {Yamada}},\ }\href {\doibase
  10.1143/JPSJ.81.011007} {\bibfield  {journal} {\bibinfo  {journal} {J. Phys.
  Soc. Jpn.}\ }\textbf {\bibinfo {volume} {81}},\ \bibinfo {pages} {011007}
  (\bibinfo {year} {2012})}\BibitemShut {NoStop}%
\bibitem [{\citenamefont {H\"ucker}\ \emph {et~al.}(2011)\citenamefont
  {H\"ucker}, \citenamefont {v.~Zimmermann}, \citenamefont {Gu}, \citenamefont
  {Xu}, \citenamefont {Wen}, \citenamefont {Xu}, \citenamefont {Kang},
  \citenamefont {Zheludev},\ and\ \citenamefont {Tranquada}}]{huck11}%
  \BibitemOpen
  \bibfield  {author} {\bibinfo {author} {\bibfnamefont {M.}~\bibnamefont
  {H\"ucker}}, \bibinfo {author} {\bibfnamefont {M.}~\bibnamefont
  {v.~Zimmermann}}, \bibinfo {author} {\bibfnamefont {G.~D.}\ \bibnamefont
  {Gu}}, \bibinfo {author} {\bibfnamefont {Z.~J.}\ \bibnamefont {Xu}}, \bibinfo
  {author} {\bibfnamefont {J.~S.}\ \bibnamefont {Wen}}, \bibinfo {author}
  {\bibfnamefont {G.}~\bibnamefont {Xu}}, \bibinfo {author} {\bibfnamefont
  {H.~J.}\ \bibnamefont {Kang}}, \bibinfo {author} {\bibfnamefont
  {A.}~\bibnamefont {Zheludev}}, \ and\ \bibinfo {author} {\bibfnamefont
  {J.~M.}\ \bibnamefont {Tranquada}},\ }\href@noop {} {\bibfield  {journal}
  {\bibinfo  {journal} {Phys. Rev. B}\ }\textbf {\bibinfo {volume} {83}},\
  \bibinfo {pages} {104506} (\bibinfo {year} {2011})}\BibitemShut {NoStop}%
\bibitem [{\citenamefont {Sasagawa}\ \emph {et~al.}(2000)\citenamefont
  {Sasagawa}, \citenamefont {Togawa}, \citenamefont {Shimoyama}, \citenamefont
  {Kapitulnik}, \citenamefont {Kitazawa},\ and\ \citenamefont
  {Kishio}}]{sasa00}%
  \BibitemOpen
  \bibfield  {author} {\bibinfo {author} {\bibfnamefont {T.}~\bibnamefont
  {Sasagawa}}, \bibinfo {author} {\bibfnamefont {Y.}~\bibnamefont {Togawa}},
  \bibinfo {author} {\bibfnamefont {J.}~\bibnamefont {Shimoyama}}, \bibinfo
  {author} {\bibfnamefont {A.}~\bibnamefont {Kapitulnik}}, \bibinfo {author}
  {\bibfnamefont {K.}~\bibnamefont {Kitazawa}}, \ and\ \bibinfo {author}
  {\bibfnamefont {K.}~\bibnamefont {Kishio}},\ }\href@noop {} {\bibfield
  {journal} {\bibinfo  {journal} {Phys. Rev. B}\ }\textbf {\bibinfo {volume}
  {61}},\ \bibinfo {pages} {1610} (\bibinfo {year} {2000})}\BibitemShut
  {NoStop}%
\bibitem [{\citenamefont {{R. Gilardi}}\ \emph {et~al.}(2005)\citenamefont {{R.
  Gilardi}}, \citenamefont {{S. Streule}}, \citenamefont {{N. Momono}},
  \citenamefont {{M. Oda}},\ and\ \citenamefont {{J. Mesot}}}]{gila05}%
  \BibitemOpen
  \bibfield  {author} {\bibinfo {author} {\bibnamefont {{R. Gilardi}}},
  \bibinfo {author} {\bibnamefont {{S. Streule}}}, \bibinfo {author}
  {\bibnamefont {{N. Momono}}}, \bibinfo {author} {\bibnamefont {{M. Oda}}}, \
  and\ \bibinfo {author} {\bibnamefont {{J. Mesot}}},\ }\href@noop {}
  {\bibfield  {journal} {\bibinfo  {journal} {Eur. Phys. J. B}\ }\textbf
  {\bibinfo {volume} {47}},\ \bibinfo {pages} {231} (\bibinfo {year}
  {2005})}\BibitemShut {NoStop}%
\bibitem [{\citenamefont {Adachi}\ \emph {et~al.}(2005)\citenamefont {Adachi},
  \citenamefont {Kitajima}, \citenamefont {Manabe}, \citenamefont {Koike},
  \citenamefont {Kudo}, \citenamefont {Sasaki},\ and\ \citenamefont
  {Kobayashi}}]{adac05}%
  \BibitemOpen
  \bibfield  {author} {\bibinfo {author} {\bibfnamefont {T.}~\bibnamefont
  {Adachi}}, \bibinfo {author} {\bibfnamefont {N.}~\bibnamefont {Kitajima}},
  \bibinfo {author} {\bibfnamefont {T.}~\bibnamefont {Manabe}}, \bibinfo
  {author} {\bibfnamefont {Y.}~\bibnamefont {Koike}}, \bibinfo {author}
  {\bibfnamefont {K.}~\bibnamefont {Kudo}}, \bibinfo {author} {\bibfnamefont
  {T.}~\bibnamefont {Sasaki}}, \ and\ \bibinfo {author} {\bibfnamefont
  {N.}~\bibnamefont {Kobayashi}},\ }\href@noop {} {\bibfield  {journal}
  {\bibinfo  {journal} {Phys. Rev. B}\ }\textbf {\bibinfo {volume} {71}},\
  \bibinfo {eid} {104516} (\bibinfo {year} {2005})}\BibitemShut {NoStop}%
\bibitem [{Note1()}]{Note1}%
  \BibitemOpen
  \bibinfo {note} {To be clear, the measurements of resistivity reported in
  Refs.~\protect \rev@citealpnum {sasa00} and \protect \rev@citealpnum {adac05}
  are nominally for $\rho _\parallel $, rather than $\rho _\bot $, but we
  expect that if they had determined $T_c(H_\bot )$ from $\rho _\bot $, they
  would have obtained results consistent with those of Ref.~\protect
  \rev@citealpnum {gila05}.}\BibitemShut {Stop}%
\bibitem [{\citenamefont {Xiang}\ \emph {et~al.}(2009)\citenamefont {Xiang},
  \citenamefont {Ding}, \citenamefont {Zhang},\ and\ \citenamefont
  {Li}}]{xian09}%
  \BibitemOpen
  \bibfield  {author} {\bibinfo {author} {\bibfnamefont {X.~Q.}\ \bibnamefont
  {Xiang}}, \bibinfo {author} {\bibfnamefont {J.~F.}\ \bibnamefont {Ding}},
  \bibinfo {author} {\bibfnamefont {Y.~Q.}\ \bibnamefont {Zhang}}, \ and\
  \bibinfo {author} {\bibfnamefont {X.~G.}\ \bibnamefont {Li}},\ }\href@noop {}
  {\bibfield  {journal} {\bibinfo  {journal} {Phys. Rev. B}\ }\textbf {\bibinfo
  {volume} {79}},\ \bibinfo {pages} {012501} (\bibinfo {year}
  {2009})}\BibitemShut {NoStop}%
\bibitem [{\citenamefont {Ambegaokar}\ and\ \citenamefont
  {Halperin}(1969)}]{ambe69}%
  \BibitemOpen
  \bibfield  {author} {\bibinfo {author} {\bibfnamefont {V.}~\bibnamefont
  {Ambegaokar}}\ and\ \bibinfo {author} {\bibfnamefont {B.~I.}\ \bibnamefont
  {Halperin}},\ }\href@noop {} {\bibfield  {journal} {\bibinfo  {journal}
  {Phys. Rev. Lett.}\ }\textbf {\bibinfo {volume} {22}},\ \bibinfo {pages}
  {1364} (\bibinfo {year} {1969})}\BibitemShut {NoStop}%
\bibitem [{\citenamefont {Brice\~no}\ \emph {et~al.}(1991)\citenamefont
  {Brice\~no}, \citenamefont {Crommie},\ and\ \citenamefont {Zettl}}]{bric91}%
  \BibitemOpen
  \bibfield  {author} {\bibinfo {author} {\bibfnamefont {G.}~\bibnamefont
  {Brice\~no}}, \bibinfo {author} {\bibfnamefont {M.~F.}\ \bibnamefont
  {Crommie}}, \ and\ \bibinfo {author} {\bibfnamefont {A.}~\bibnamefont
  {Zettl}},\ }\href@noop {} {\bibfield  {journal} {\bibinfo  {journal} {Phys.
  Rev. Lett.}\ }\textbf {\bibinfo {volume} {66}},\ \bibinfo {pages} {2164}
  (\bibinfo {year} {1991})}\BibitemShut {NoStop}%
\bibitem [{\citenamefont {Gray}\ and\ \citenamefont {Kim}(1993)}]{gray93}%
  \BibitemOpen
  \bibfield  {author} {\bibinfo {author} {\bibfnamefont {K.~E.}\ \bibnamefont
  {Gray}}\ and\ \bibinfo {author} {\bibfnamefont {D.~H.}\ \bibnamefont {Kim}},\
  }\href@noop {} {\bibfield  {journal} {\bibinfo  {journal} {Phys. Rev. Lett.}\
  }\textbf {\bibinfo {volume} {70}},\ \bibinfo {pages} {1693} (\bibinfo {year}
  {1993})}\BibitemShut {NoStop}%
\bibitem [{\citenamefont {Hettinger}\ \emph {et~al.}(1995)\citenamefont
  {Hettinger}, \citenamefont {Gray}, \citenamefont {Veal}, \citenamefont
  {Paulikas}, \citenamefont {Kostic}, \citenamefont {Washburn}, \citenamefont
  {Tonjes},\ and\ \citenamefont {Flewelling}}]{hett95}%
  \BibitemOpen
  \bibfield  {author} {\bibinfo {author} {\bibfnamefont {J.~D.}\ \bibnamefont
  {Hettinger}}, \bibinfo {author} {\bibfnamefont {K.~E.}\ \bibnamefont {Gray}},
  \bibinfo {author} {\bibfnamefont {B.~W.}\ \bibnamefont {Veal}}, \bibinfo
  {author} {\bibfnamefont {A.~P.}\ \bibnamefont {Paulikas}}, \bibinfo {author}
  {\bibfnamefont {P.}~\bibnamefont {Kostic}}, \bibinfo {author} {\bibfnamefont
  {B.~R.}\ \bibnamefont {Washburn}}, \bibinfo {author} {\bibfnamefont {W.~C.}\
  \bibnamefont {Tonjes}}, \ and\ \bibinfo {author} {\bibfnamefont {A.~C.}\
  \bibnamefont {Flewelling}},\ }\href {\doibase 10.1103/PhysRevLett.74.4726}
  {\bibfield  {journal} {\bibinfo  {journal} {Phys. Rev. Lett.}\ }\textbf
  {\bibinfo {volume} {74}},\ \bibinfo {pages} {4726} (\bibinfo {year}
  {1995})}\BibitemShut {NoStop}%
\bibitem [{\citenamefont {Suzuki}\ \emph {et~al.}(1998)\citenamefont {Suzuki},
  \citenamefont {Watanabe},\ and\ \citenamefont {Matsuda}}]{suzu98b}%
  \BibitemOpen
  \bibfield  {author} {\bibinfo {author} {\bibfnamefont {M.}~\bibnamefont
  {Suzuki}}, \bibinfo {author} {\bibfnamefont {T.}~\bibnamefont {Watanabe}}, \
  and\ \bibinfo {author} {\bibfnamefont {A.}~\bibnamefont {Matsuda}},\
  }\href@noop {} {\bibfield  {journal} {\bibinfo  {journal} {Phys. Rev. Lett.}\
  }\textbf {\bibinfo {volume} {81}},\ \bibinfo {pages} {4248} (\bibinfo {year}
  {1998})}\BibitemShut {NoStop}%
\bibitem [{\citenamefont {Axe}\ and\ \citenamefont {Crawford}(1994)}]{axe94}%
  \BibitemOpen
  \bibfield  {author} {\bibinfo {author} {\bibfnamefont {J.~D.}\ \bibnamefont
  {Axe}}\ and\ \bibinfo {author} {\bibfnamefont {M.~K.}\ \bibnamefont
  {Crawford}},\ }\href@noop {} {\bibfield  {journal} {\bibinfo  {journal} {J.
  Low Temp. Phys.}\ }\textbf {\bibinfo {volume} {95}},\ \bibinfo {pages} {271}
  (\bibinfo {year} {1994})}\BibitemShut {NoStop}%
\bibitem [{\citenamefont {Wen}\ \emph {et~al.}()\citenamefont {Wen},
  \citenamefont {Xu}, \citenamefont {Xu}, \citenamefont {Jie}, \citenamefont
  {H{\"u}cker}, \citenamefont {Zheludev}, \citenamefont {Tian}, \citenamefont
  {Winn}, \citenamefont {Zarestky}, \citenamefont {Singh}, \citenamefont
  {Hong}, \citenamefont {Li}, \citenamefont {Gu},\ and\ \citenamefont
  {Tranquada}}]{wen12a}%
  \BibitemOpen
  \bibfield  {author} {\bibinfo {author} {\bibfnamefont {J.}~\bibnamefont
  {Wen}}, \bibinfo {author} {\bibfnamefont {Z.}~\bibnamefont {Xu}}, \bibinfo
  {author} {\bibfnamefont {G.}~\bibnamefont {Xu}}, \bibinfo {author}
  {\bibfnamefont {Q.}~\bibnamefont {Jie}}, \bibinfo {author} {\bibfnamefont
  {M.}~\bibnamefont {H{\"u}cker}}, \bibinfo {author} {\bibfnamefont
  {A.}~\bibnamefont {Zheludev}}, \bibinfo {author} {\bibfnamefont
  {W.}~\bibnamefont {Tian}}, \bibinfo {author} {\bibfnamefont {B.~L.}\
  \bibnamefont {Winn}}, \bibinfo {author} {\bibfnamefont {J.~L.}\ \bibnamefont
  {Zarestky}}, \bibinfo {author} {\bibfnamefont {D.~K.}\ \bibnamefont {Singh}},
  \bibinfo {author} {\bibfnamefont {T.}~\bibnamefont {Hong}}, \bibinfo {author}
  {\bibfnamefont {Q.}~\bibnamefont {Li}}, \bibinfo {author} {\bibfnamefont
  {G.}~\bibnamefont {Gu}}, \ and\ \bibinfo {author} {\bibfnamefont {J.~M.}\
  \bibnamefont {Tranquada}},\ }\href@noop {} {\enquote {\bibinfo {title}
  {{Probing the connections between superconductivity, stripe order, and
  structure in La$_{1.905}$Ba$_{0.095}$CuO$_4$}},}\ }\Eprint
  {http://arxiv.org/abs/arXiv:1111.5383} {arXiv:1111.5383} \BibitemShut
  {NoStop}%
\bibitem [{\citenamefont {Homes}\ \emph {et~al.}()\citenamefont {Homes},
  \citenamefont {{H\"ucker}}, \citenamefont {Li}, \citenamefont {Xu},
  \citenamefont {Wen}, \citenamefont {Gu},\ and\ \citenamefont
  {Tranquada}}]{home12}%
  \BibitemOpen
  \bibfield  {author} {\bibinfo {author} {\bibfnamefont {C.~C.}\ \bibnamefont
  {Homes}}, \bibinfo {author} {\bibfnamefont {M.}~\bibnamefont {{H\"ucker}}},
  \bibinfo {author} {\bibfnamefont {Q.}~\bibnamefont {Li}}, \bibinfo {author}
  {\bibfnamefont {Z.~J.}\ \bibnamefont {Xu}}, \bibinfo {author} {\bibfnamefont
  {J.~S.}\ \bibnamefont {Wen}}, \bibinfo {author} {\bibfnamefont {G.~D.}\
  \bibnamefont {Gu}}, \ and\ \bibinfo {author} {\bibfnamefont {J.~M.}\
  \bibnamefont {Tranquada}},\ }\href@noop {} {}\bibinfo {howpublished} {Phys.
  Rev. B (accepted)},\ \Eprint {http://arxiv.org/abs/arXiv:1110.2118}
  {arXiv:1110.2118} \BibitemShut {NoStop}%
\bibitem [{\citenamefont {Tranquada}\ \emph {et~al.}(1995)\citenamefont
  {Tranquada}, \citenamefont {Sternlieb}, \citenamefont {Axe}, \citenamefont
  {Nakamura},\ and\ \citenamefont {Uchida}}]{tran95a}%
  \BibitemOpen
  \bibfield  {author} {\bibinfo {author} {\bibfnamefont {J.~M.}\ \bibnamefont
  {Tranquada}}, \bibinfo {author} {\bibfnamefont {B.~J.}\ \bibnamefont
  {Sternlieb}}, \bibinfo {author} {\bibfnamefont {J.~D.}\ \bibnamefont {Axe}},
  \bibinfo {author} {\bibfnamefont {Y.}~\bibnamefont {Nakamura}}, \ and\
  \bibinfo {author} {\bibfnamefont {S.}~\bibnamefont {Uchida}},\ }\href@noop {}
  {\bibfield  {journal} {\bibinfo  {journal} {Nature}\ }\textbf {\bibinfo
  {volume} {375}},\ \bibinfo {pages} {561} (\bibinfo {year}
  {1995})}\BibitemShut {NoStop}%
\bibitem [{\citenamefont {Fujita}\ \emph {et~al.}(2004)\citenamefont {Fujita},
  \citenamefont {Goka}, \citenamefont {Yamada}, \citenamefont {Tranquada},\
  and\ \citenamefont {Regnault}}]{fuji04}%
  \BibitemOpen
  \bibfield  {author} {\bibinfo {author} {\bibfnamefont {M.}~\bibnamefont
  {Fujita}}, \bibinfo {author} {\bibfnamefont {H.}~\bibnamefont {Goka}},
  \bibinfo {author} {\bibfnamefont {K.}~\bibnamefont {Yamada}}, \bibinfo
  {author} {\bibfnamefont {J.~M.}\ \bibnamefont {Tranquada}}, \ and\ \bibinfo
  {author} {\bibfnamefont {L.~P.}\ \bibnamefont {Regnault}},\ }\href@noop {}
  {\bibfield  {journal} {\bibinfo  {journal} {Phys. Rev. B}\ }\textbf {\bibinfo
  {volume} {70}},\ \bibinfo {pages} {104517} (\bibinfo {year}
  {2004})}\BibitemShut {NoStop}%
\bibitem [{\citenamefont {Li}\ \emph {et~al.}(2007)\citenamefont {Li},
  \citenamefont {{H\"ucker}}, \citenamefont {Gu}, \citenamefont {Tsvelik},\
  and\ \citenamefont {Tranquada}}]{li07}%
  \BibitemOpen
  \bibfield  {author} {\bibinfo {author} {\bibfnamefont {Q.}~\bibnamefont
  {Li}}, \bibinfo {author} {\bibfnamefont {M.}~\bibnamefont {{H\"ucker}}},
  \bibinfo {author} {\bibfnamefont {G.~D.}\ \bibnamefont {Gu}}, \bibinfo
  {author} {\bibfnamefont {A.~M.}\ \bibnamefont {Tsvelik}}, \ and\ \bibinfo
  {author} {\bibfnamefont {J.~M.}\ \bibnamefont {Tranquada}},\ }\href@noop {}
  {\bibfield  {journal} {\bibinfo  {journal} {Phys. Rev. Lett.}\ }\textbf
  {\bibinfo {volume} {99}},\ \bibinfo {eid} {067001} (\bibinfo {year}
  {2007})}\BibitemShut {NoStop}%
\bibitem [{\citenamefont {Tranquada}\ \emph {et~al.}(2008)\citenamefont
  {Tranquada}, \citenamefont {Gu}, \citenamefont {H{\"u}cker}, \citenamefont
  {Jie}, \citenamefont {Kang}, \citenamefont {Klingeler}, \citenamefont {Li},
  \citenamefont {Tristan}, \citenamefont {Wen}, \citenamefont {Xu},
  \citenamefont {Xu}, \citenamefont {Zhou},\ and\ \citenamefont
  {v.~Zimmermann}}]{tran08}%
  \BibitemOpen
  \bibfield  {author} {\bibinfo {author} {\bibfnamefont {J.~M.}\ \bibnamefont
  {Tranquada}}, \bibinfo {author} {\bibfnamefont {G.~D.}\ \bibnamefont {Gu}},
  \bibinfo {author} {\bibfnamefont {M.}~\bibnamefont {H{\"u}cker}}, \bibinfo
  {author} {\bibfnamefont {Q.}~\bibnamefont {Jie}}, \bibinfo {author}
  {\bibfnamefont {H.-J.}\ \bibnamefont {Kang}}, \bibinfo {author}
  {\bibfnamefont {R.}~\bibnamefont {Klingeler}}, \bibinfo {author}
  {\bibfnamefont {Q.}~\bibnamefont {Li}}, \bibinfo {author} {\bibfnamefont
  {N.}~\bibnamefont {Tristan}}, \bibinfo {author} {\bibfnamefont {J.~S.}\
  \bibnamefont {Wen}}, \bibinfo {author} {\bibfnamefont {G.~Y.}\ \bibnamefont
  {Xu}}, \bibinfo {author} {\bibfnamefont {Z.~J.}\ \bibnamefont {Xu}}, \bibinfo
  {author} {\bibfnamefont {J.}~\bibnamefont {Zhou}}, \ and\ \bibinfo {author}
  {\bibfnamefont {M.}~\bibnamefont {v.~Zimmermann}},\ }\href@noop {} {\bibfield
   {journal} {\bibinfo  {journal} {Phys. Rev. B}\ }\textbf {\bibinfo {volume}
  {78}},\ \bibinfo {eid} {174529} (\bibinfo {year} {2008})}\BibitemShut
  {NoStop}%
\bibitem [{\citenamefont {Himeda}\ \emph {et~al.}(2002)\citenamefont {Himeda},
  \citenamefont {Kato},\ and\ \citenamefont {Ogata}}]{hime02}%
  \BibitemOpen
  \bibfield  {author} {\bibinfo {author} {\bibfnamefont {A.}~\bibnamefont
  {Himeda}}, \bibinfo {author} {\bibfnamefont {T.}~\bibnamefont {Kato}}, \ and\
  \bibinfo {author} {\bibfnamefont {M.}~\bibnamefont {Ogata}},\ }\href@noop {}
  {\bibfield  {journal} {\bibinfo  {journal} {Phys. Rev. Lett.}\ }\textbf
  {\bibinfo {volume} {88}},\ \bibinfo {pages} {117001} (\bibinfo {year}
  {2002})}\BibitemShut {NoStop}%
\bibitem [{\citenamefont {Berg}\ \emph {et~al.}(2007)\citenamefont {Berg},
  \citenamefont {Fradkin}, \citenamefont {Kim}, \citenamefont {Kivelson},
  \citenamefont {Oganesyan}, \citenamefont {Tranquada},\ and\ \citenamefont
  {Zhang}}]{berg07}%
  \BibitemOpen
  \bibfield  {author} {\bibinfo {author} {\bibfnamefont {E.}~\bibnamefont
  {Berg}}, \bibinfo {author} {\bibfnamefont {E.}~\bibnamefont {Fradkin}},
  \bibinfo {author} {\bibfnamefont {E.-A.}\ \bibnamefont {Kim}}, \bibinfo
  {author} {\bibfnamefont {S.~A.}\ \bibnamefont {Kivelson}}, \bibinfo {author}
  {\bibfnamefont {V.}~\bibnamefont {Oganesyan}}, \bibinfo {author}
  {\bibfnamefont {J.~M.}\ \bibnamefont {Tranquada}}, \ and\ \bibinfo {author}
  {\bibfnamefont {S.~C.}\ \bibnamefont {Zhang}},\ }\href@noop {} {\bibfield
  {journal} {\bibinfo  {journal} {Phys. Rev. Lett.}\ }\textbf {\bibinfo
  {volume} {99}},\ \bibinfo {eid} {127003} (\bibinfo {year}
  {2007})}\BibitemShut {NoStop}%
\bibitem [{\citenamefont {Fuchs}\ \emph {et~al.}(1998)\citenamefont {Fuchs},
  \citenamefont {Zeldov}, \citenamefont {Rappaport}, \citenamefont {Tamegai},
  \citenamefont {Ooi},\ and\ \citenamefont {Shtrikman}}]{fuch98}%
  \BibitemOpen
  \bibfield  {author} {\bibinfo {author} {\bibfnamefont {D.~T.}\ \bibnamefont
  {Fuchs}}, \bibinfo {author} {\bibfnamefont {E.}~\bibnamefont {Zeldov}},
  \bibinfo {author} {\bibfnamefont {M.}~\bibnamefont {Rappaport}}, \bibinfo
  {author} {\bibfnamefont {T.}~\bibnamefont {Tamegai}}, \bibinfo {author}
  {\bibfnamefont {S.}~\bibnamefont {Ooi}}, \ and\ \bibinfo {author}
  {\bibfnamefont {H.}~\bibnamefont {Shtrikman}},\ }\href@noop {} {\bibfield
  {journal} {\bibinfo  {journal} {Nature}\ }\textbf {\bibinfo {volume} {391}},\
  \bibinfo {pages} {373} (\bibinfo {year} {1998})}\BibitemShut {NoStop}%
\bibitem [{\citenamefont {Beidenkopf}\ \emph {et~al.}(2009)\citenamefont
  {Beidenkopf}, \citenamefont {Myasoedov}, \citenamefont {Zeldov},
  \citenamefont {Brandt}, \citenamefont {Mikitik}, \citenamefont {Tamegai},
  \citenamefont {Sasagawa},\ and\ \citenamefont {van~der Beek}}]{beid09}%
  \BibitemOpen
  \bibfield  {author} {\bibinfo {author} {\bibfnamefont {H.}~\bibnamefont
  {Beidenkopf}}, \bibinfo {author} {\bibfnamefont {Y.}~\bibnamefont
  {Myasoedov}}, \bibinfo {author} {\bibfnamefont {E.}~\bibnamefont {Zeldov}},
  \bibinfo {author} {\bibfnamefont {E.~H.}\ \bibnamefont {Brandt}}, \bibinfo
  {author} {\bibfnamefont {G.~P.}\ \bibnamefont {Mikitik}}, \bibinfo {author}
  {\bibfnamefont {T.}~\bibnamefont {Tamegai}}, \bibinfo {author} {\bibfnamefont
  {T.}~\bibnamefont {Sasagawa}}, \ and\ \bibinfo {author} {\bibfnamefont
  {C.~J.}\ \bibnamefont {van~der Beek}},\ }\href@noop {} {\bibfield  {journal}
  {\bibinfo  {journal} {Phys. Rev. B}\ }\textbf {\bibinfo {volume} {80}},\
  \bibinfo {pages} {224526} (\bibinfo {year} {2009})}\BibitemShut {NoStop}%
\bibitem [{\citenamefont {Busch}\ \emph {et~al.}(1992)\citenamefont {Busch},
  \citenamefont {Ries}, \citenamefont {Werthner}, \citenamefont
  {Kreiselmeyer},\ and\ \citenamefont {Saemann-Ischenko}}]{busc92}%
  \BibitemOpen
  \bibfield  {author} {\bibinfo {author} {\bibfnamefont {R.}~\bibnamefont
  {Busch}}, \bibinfo {author} {\bibfnamefont {G.}~\bibnamefont {Ries}},
  \bibinfo {author} {\bibfnamefont {H.}~\bibnamefont {Werthner}}, \bibinfo
  {author} {\bibfnamefont {G.}~\bibnamefont {Kreiselmeyer}}, \ and\ \bibinfo
  {author} {\bibfnamefont {G.}~\bibnamefont {Saemann-Ischenko}},\ }\href
  {\doibase 10.1103/PhysRevLett.69.522} {\bibfield  {journal} {\bibinfo
  {journal} {Phys. Rev. Lett.}\ }\textbf {\bibinfo {volume} {69}},\ \bibinfo
  {pages} {522} (\bibinfo {year} {1992})}\BibitemShut {NoStop}%
\bibitem [{\citenamefont {Cho}\ \emph {et~al.}(1994)\citenamefont {Cho},
  \citenamefont {Maley}, \citenamefont {Fleshler}, \citenamefont {Lacerda},\
  and\ \citenamefont {Bulaevskii}}]{cho94}%
  \BibitemOpen
  \bibfield  {author} {\bibinfo {author} {\bibfnamefont {J.~H.}\ \bibnamefont
  {Cho}}, \bibinfo {author} {\bibfnamefont {M.~P.}\ \bibnamefont {Maley}},
  \bibinfo {author} {\bibfnamefont {S.}~\bibnamefont {Fleshler}}, \bibinfo
  {author} {\bibfnamefont {A.}~\bibnamefont {Lacerda}}, \ and\ \bibinfo
  {author} {\bibfnamefont {L.~N.}\ \bibnamefont {Bulaevskii}},\ }\href@noop {}
  {\bibfield  {journal} {\bibinfo  {journal} {Phys. Rev. B}\ }\textbf {\bibinfo
  {volume} {50}},\ \bibinfo {pages} {6493} (\bibinfo {year}
  {1994})}\BibitemShut {NoStop}%
\bibitem [{\citenamefont {Reehuis}\ \emph {et~al.}(2006)\citenamefont
  {Reehuis}, \citenamefont {Ulrich}, \citenamefont {Proke\v{s}}, \citenamefont
  {Gozar}, \citenamefont {Blumberg}, \citenamefont {Komiya}, \citenamefont
  {Ando}, \citenamefont {Pattison},\ and\ \citenamefont {Keimer}}]{reeh06}%
  \BibitemOpen
  \bibfield  {author} {\bibinfo {author} {\bibfnamefont {M.}~\bibnamefont
  {Reehuis}}, \bibinfo {author} {\bibfnamefont {C.}~\bibnamefont {Ulrich}},
  \bibinfo {author} {\bibfnamefont {K.}~\bibnamefont {Proke\v{s}}}, \bibinfo
  {author} {\bibfnamefont {A.}~\bibnamefont {Gozar}}, \bibinfo {author}
  {\bibfnamefont {G.}~\bibnamefont {Blumberg}}, \bibinfo {author}
  {\bibfnamefont {S.}~\bibnamefont {Komiya}}, \bibinfo {author} {\bibfnamefont
  {Y.}~\bibnamefont {Ando}}, \bibinfo {author} {\bibfnamefont {P.}~\bibnamefont
  {Pattison}}, \ and\ \bibinfo {author} {\bibfnamefont {B.}~\bibnamefont
  {Keimer}},\ }\href {\doibase 10.1103/PhysRevB.73.144513} {\bibfield
  {journal} {\bibinfo  {journal} {Phys. Rev. B}\ }\textbf {\bibinfo {volume}
  {73}},\ \bibinfo {pages} {144513} (\bibinfo {year} {2006})}\BibitemShut
  {NoStop}%
\bibitem [{\citenamefont {Sasagawa}\ \emph {et~al.}(1998)\citenamefont
  {Sasagawa}, \citenamefont {Kishio}, \citenamefont {Togawa}, \citenamefont
  {Shimoyama},\ and\ \citenamefont {Kitazawa}}]{sasa98}%
  \BibitemOpen
  \bibfield  {author} {\bibinfo {author} {\bibfnamefont {T.}~\bibnamefont
  {Sasagawa}}, \bibinfo {author} {\bibfnamefont {K.}~\bibnamefont {Kishio}},
  \bibinfo {author} {\bibfnamefont {Y.}~\bibnamefont {Togawa}}, \bibinfo
  {author} {\bibfnamefont {J.}~\bibnamefont {Shimoyama}}, \ and\ \bibinfo
  {author} {\bibfnamefont {K.}~\bibnamefont {Kitazawa}},\ }\href@noop {}
  {\bibfield  {journal} {\bibinfo  {journal} {Phys. Rev. Lett.}\ }\textbf
  {\bibinfo {volume} {80}},\ \bibinfo {pages} {4297} (\bibinfo {year}
  {1998})}\BibitemShut {NoStop}%
\bibitem [{\citenamefont {Ando}\ \emph {et~al.}(1999)\citenamefont {Ando},
  \citenamefont {Boebinger}, \citenamefont {Passner}, \citenamefont
  {Schneemeyer}, \citenamefont {Kimura}, \citenamefont {Okuya}, \citenamefont
  {Watauchi}, \citenamefont {Shimoyama}, \citenamefont {Kishio}, \citenamefont
  {Tamasaku}, \citenamefont {Ichikawa},\ and\ \citenamefont
  {Uchida}}]{ando99b}%
  \BibitemOpen
  \bibfield  {author} {\bibinfo {author} {\bibfnamefont {Y.}~\bibnamefont
  {Ando}}, \bibinfo {author} {\bibfnamefont {G.~S.}\ \bibnamefont {Boebinger}},
  \bibinfo {author} {\bibfnamefont {A.}~\bibnamefont {Passner}}, \bibinfo
  {author} {\bibfnamefont {L.~F.}\ \bibnamefont {Schneemeyer}}, \bibinfo
  {author} {\bibfnamefont {T.}~\bibnamefont {Kimura}}, \bibinfo {author}
  {\bibfnamefont {M.}~\bibnamefont {Okuya}}, \bibinfo {author} {\bibfnamefont
  {S.}~\bibnamefont {Watauchi}}, \bibinfo {author} {\bibfnamefont
  {J.}~\bibnamefont {Shimoyama}}, \bibinfo {author} {\bibfnamefont
  {K.}~\bibnamefont {Kishio}}, \bibinfo {author} {\bibfnamefont
  {K.}~\bibnamefont {Tamasaku}}, \bibinfo {author} {\bibfnamefont
  {N.}~\bibnamefont {Ichikawa}}, \ and\ \bibinfo {author} {\bibfnamefont
  {S.}~\bibnamefont {Uchida}},\ }\href@noop {} {\bibfield  {journal} {\bibinfo
  {journal} {Phys. Rev. B}\ }\textbf {\bibinfo {volume} {60}},\ \bibinfo
  {pages} {12475} (\bibinfo {year} {1999})}\BibitemShut {NoStop}%
\bibitem [{\citenamefont {{H\"ucker}}\ \emph {et~al.}(2008)\citenamefont
  {{H\"ucker}}, \citenamefont {Gu},\ and\ \citenamefont {Tranquada}}]{huck08}%
  \BibitemOpen
  \bibfield  {author} {\bibinfo {author} {\bibfnamefont {M.}~\bibnamefont
  {{H\"ucker}}}, \bibinfo {author} {\bibfnamefont {G.~D.}\ \bibnamefont {Gu}},
  \ and\ \bibinfo {author} {\bibfnamefont {J.~M.}\ \bibnamefont {Tranquada}},\
  }\href@noop {} {\bibfield  {journal} {\bibinfo  {journal} {Phys. Rev. B}\
  }\textbf {\bibinfo {volume} {78}},\ \bibinfo {eid} {214507} (\bibinfo {year}
  {2008})}\BibitemShut {NoStop}%
\bibitem [{\citenamefont {Tinkham}(1988)}]{tink88}%
  \BibitemOpen
  \bibfield  {author} {\bibinfo {author} {\bibfnamefont {M.}~\bibnamefont
  {Tinkham}},\ }\href@noop {} {\bibfield  {journal} {\bibinfo  {journal} {Phys.
  Rev. Lett.}\ }\textbf {\bibinfo {volume} {61}},\ \bibinfo {pages} {1658}
  (\bibinfo {year} {1988})}\BibitemShut {NoStop}%
\bibitem [{\citenamefont {McCumber}(1968)}]{mccu68}%
  \BibitemOpen
  \bibfield  {author} {\bibinfo {author} {\bibfnamefont {D.~E.}\ \bibnamefont
  {McCumber}},\ }\href@noop {} {\bibfield  {journal} {\bibinfo  {journal} {J.
  Appl. Phys.}\ }\textbf {\bibinfo {volume} {39}},\ \bibinfo {pages} {3113}
  (\bibinfo {year} {1968})}\BibitemShut {NoStop}%
\bibitem [{\citenamefont {Latyshev}\ \emph {et~al.}(1999)\citenamefont
  {Latyshev}, \citenamefont {Yamashita}, \citenamefont {Bulaevskii},
  \citenamefont {Graf}, \citenamefont {Balatsky},\ and\ \citenamefont
  {Maley}}]{laty99}%
  \BibitemOpen
  \bibfield  {author} {\bibinfo {author} {\bibfnamefont {Y.~I.}\ \bibnamefont
  {Latyshev}}, \bibinfo {author} {\bibfnamefont {T.}~\bibnamefont {Yamashita}},
  \bibinfo {author} {\bibfnamefont {L.~N.}\ \bibnamefont {Bulaevskii}},
  \bibinfo {author} {\bibfnamefont {M.~J.}\ \bibnamefont {Graf}}, \bibinfo
  {author} {\bibfnamefont {A.~V.}\ \bibnamefont {Balatsky}}, \ and\ \bibinfo
  {author} {\bibfnamefont {M.~P.}\ \bibnamefont {Maley}},\ }\href {\doibase
  10.1103/PhysRevLett.82.5345} {\bibfield  {journal} {\bibinfo  {journal}
  {Phys. Rev. Lett.}\ }\textbf {\bibinfo {volume} {82}},\ \bibinfo {pages}
  {5345} (\bibinfo {year} {1999})}\BibitemShut {NoStop}%
\bibitem [{\citenamefont {H\"ucker}\ \emph {et~al.}(2010)\citenamefont
  {H\"ucker}, \citenamefont {v.~Zimmermann}, \citenamefont {Debessai},
  \citenamefont {Schilling}, \citenamefont {Tranquada},\ and\ \citenamefont
  {Gu}}]{huck10}%
  \BibitemOpen
  \bibfield  {author} {\bibinfo {author} {\bibfnamefont {M.}~\bibnamefont
  {H\"ucker}}, \bibinfo {author} {\bibfnamefont {M.}~\bibnamefont
  {v.~Zimmermann}}, \bibinfo {author} {\bibfnamefont {M.}~\bibnamefont
  {Debessai}}, \bibinfo {author} {\bibfnamefont {J.~S.}\ \bibnamefont
  {Schilling}}, \bibinfo {author} {\bibfnamefont {J.~M.}\ \bibnamefont
  {Tranquada}}, \ and\ \bibinfo {author} {\bibfnamefont {G.~D.}\ \bibnamefont
  {Gu}},\ }\href@noop {} {\bibfield  {journal} {\bibinfo  {journal} {Phys. Rev.
  Lett.}\ }\textbf {\bibinfo {volume} {104}},\ \bibinfo {pages} {057004}
  (\bibinfo {year} {2010})}\BibitemShut {NoStop}%
\bibitem [{\citenamefont {Lake}\ \emph {et~al.}(2001)\citenamefont {Lake},
  \citenamefont {Aeppli}, \citenamefont {Clausen}, \citenamefont {McMorrow},
  \citenamefont {Lefmann}, \citenamefont {Hussey}, \citenamefont
  {Mangkorntong}, \citenamefont {Nohara}, \citenamefont {Takagi}, \citenamefont
  {Mason},\ and\ \citenamefont {Schroder}}]{lake01}%
  \BibitemOpen
  \bibfield  {author} {\bibinfo {author} {\bibfnamefont {B.}~\bibnamefont
  {Lake}}, \bibinfo {author} {\bibfnamefont {G.}~\bibnamefont {Aeppli}},
  \bibinfo {author} {\bibfnamefont {K.~N.}\ \bibnamefont {Clausen}}, \bibinfo
  {author} {\bibfnamefont {D.~F.}\ \bibnamefont {McMorrow}}, \bibinfo {author}
  {\bibfnamefont {K.}~\bibnamefont {Lefmann}}, \bibinfo {author} {\bibfnamefont
  {N.~E.}\ \bibnamefont {Hussey}}, \bibinfo {author} {\bibfnamefont
  {N.}~\bibnamefont {Mangkorntong}}, \bibinfo {author} {\bibfnamefont
  {M.}~\bibnamefont {Nohara}}, \bibinfo {author} {\bibfnamefont
  {H.}~\bibnamefont {Takagi}}, \bibinfo {author} {\bibfnamefont {T.~E.}\
  \bibnamefont {Mason}}, \ and\ \bibinfo {author} {\bibfnamefont
  {A.}~\bibnamefont {Schroder}},\ }\href@noop {} {\bibfield  {journal}
  {\bibinfo  {journal} {Science}\ }\textbf {\bibinfo {volume} {291}},\ \bibinfo
  {pages} {1759} (\bibinfo {year} {2001})}\BibitemShut {NoStop}%
\bibitem [{\citenamefont {Hoffman}\ \emph {et~al.}(2002)\citenamefont
  {Hoffman}, \citenamefont {Hudson}, \citenamefont {Lang}, \citenamefont
  {Madhavan}, \citenamefont {Eisaki}, \citenamefont {Uchida},\ and\
  \citenamefont {Davis}}]{hoff02}%
  \BibitemOpen
  \bibfield  {author} {\bibinfo {author} {\bibfnamefont {J.~E.}\ \bibnamefont
  {Hoffman}}, \bibinfo {author} {\bibfnamefont {E.~W.}\ \bibnamefont {Hudson}},
  \bibinfo {author} {\bibfnamefont {K.~M.}\ \bibnamefont {Lang}}, \bibinfo
  {author} {\bibfnamefont {V.}~\bibnamefont {Madhavan}}, \bibinfo {author}
  {\bibfnamefont {H.}~\bibnamefont {Eisaki}}, \bibinfo {author} {\bibfnamefont
  {S.}~\bibnamefont {Uchida}}, \ and\ \bibinfo {author} {\bibfnamefont {J.~C.}\
  \bibnamefont {Davis}},\ }\href@noop {} {\bibfield  {journal} {\bibinfo
  {journal} {Science}\ }\textbf {\bibinfo {volume} {295}},\ \bibinfo {pages}
  {466} (\bibinfo {year} {2002})}\BibitemShut {NoStop}%
\bibitem [{\citenamefont {Kivelson}\ \emph {et~al.}(2002)\citenamefont
  {Kivelson}, \citenamefont {Lee}, \citenamefont {Fradkin},\ and\ \citenamefont
  {Oganesyan}}]{kive02}%
  \BibitemOpen
  \bibfield  {author} {\bibinfo {author} {\bibfnamefont {S.~A.}\ \bibnamefont
  {Kivelson}}, \bibinfo {author} {\bibfnamefont {D.-H.}\ \bibnamefont {Lee}},
  \bibinfo {author} {\bibfnamefont {E.}~\bibnamefont {Fradkin}}, \ and\
  \bibinfo {author} {\bibfnamefont {V.}~\bibnamefont {Oganesyan}},\ }\href@noop
  {} {\bibfield  {journal} {\bibinfo  {journal} {Phys. Rev. B}\ }\textbf
  {\bibinfo {volume} {66}},\ \bibinfo {pages} {144516} (\bibinfo {year}
  {2002})}\BibitemShut {NoStop}%
\bibitem [{\citenamefont {Dunsiger}\ \emph {et~al.}(2008)\citenamefont
  {Dunsiger}, \citenamefont {Zhao}, \citenamefont {Yamani}, \citenamefont
  {Buyers}, \citenamefont {Dabkowska},\ and\ \citenamefont {Gaulin}}]{duns08}%
  \BibitemOpen
  \bibfield  {author} {\bibinfo {author} {\bibfnamefont {S.~R.}\ \bibnamefont
  {Dunsiger}}, \bibinfo {author} {\bibfnamefont {Y.}~\bibnamefont {Zhao}},
  \bibinfo {author} {\bibfnamefont {Z.}~\bibnamefont {Yamani}}, \bibinfo
  {author} {\bibfnamefont {W.~J.~L.}\ \bibnamefont {Buyers}}, \bibinfo {author}
  {\bibfnamefont {H.}~\bibnamefont {Dabkowska}}, \ and\ \bibinfo {author}
  {\bibfnamefont {B.~D.}\ \bibnamefont {Gaulin}},\ }\href@noop {} {\bibfield
  {journal} {\bibinfo  {journal} {Phys. Rev. B}\ }\textbf {\bibinfo {volume}
  {77}},\ \bibinfo {pages} {224410} (\bibinfo {year} {2008})}\BibitemShut
  {NoStop}%
\bibitem [{\citenamefont {Raman}\ \emph {et~al.}(2009)\citenamefont {Raman},
  \citenamefont {Oganesyan},\ and\ \citenamefont {Sondhi}}]{rama09}%
  \BibitemOpen
  \bibfield  {author} {\bibinfo {author} {\bibfnamefont {K.~S.}\ \bibnamefont
  {Raman}}, \bibinfo {author} {\bibfnamefont {V.}~\bibnamefont {Oganesyan}}, \
  and\ \bibinfo {author} {\bibfnamefont {S.~L.}\ \bibnamefont {Sondhi}},\
  }\href@noop {} {\bibfield  {journal} {\bibinfo  {journal} {Phys. Rev. B}\
  }\textbf {\bibinfo {volume} {79}},\ \bibinfo {pages} {174528} (\bibinfo
  {year} {2009})}\BibitemShut {NoStop}%
\bibitem [{\citenamefont {Berg}\ \emph {et~al.}(2009)\citenamefont {Berg},
  \citenamefont {Fradkin}, \citenamefont {Kivelson},\ and\ \citenamefont
  {Tranquada}}]{berg09b}%
  \BibitemOpen
  \bibfield  {author} {\bibinfo {author} {\bibfnamefont {E.}~\bibnamefont
  {Berg}}, \bibinfo {author} {\bibfnamefont {E.}~\bibnamefont {Fradkin}},
  \bibinfo {author} {\bibfnamefont {S.~A.}\ \bibnamefont {Kivelson}}, \ and\
  \bibinfo {author} {\bibfnamefont {J.~M.}\ \bibnamefont {Tranquada}},\
  }\href@noop {} {\bibfield  {journal} {\bibinfo  {journal} {New J. Phys.}\
  }\textbf {\bibinfo {volume} {11}},\ \bibinfo {pages} {115004} (\bibinfo
  {year} {2009})}\BibitemShut {NoStop}%
\bibitem [{\citenamefont {Tajima}\ \emph {et~al.}(2001)\citenamefont {Tajima},
  \citenamefont {Noda}, \citenamefont {Eisaki},\ and\ \citenamefont
  {Uchida}}]{taji01}%
  \BibitemOpen
  \bibfield  {author} {\bibinfo {author} {\bibfnamefont {S.}~\bibnamefont
  {Tajima}}, \bibinfo {author} {\bibfnamefont {T.}~\bibnamefont {Noda}},
  \bibinfo {author} {\bibfnamefont {H.}~\bibnamefont {Eisaki}}, \ and\ \bibinfo
  {author} {\bibfnamefont {S.}~\bibnamefont {Uchida}},\ }\href@noop {}
  {\bibfield  {journal} {\bibinfo  {journal} {Phys. Rev. Lett.}\ }\textbf
  {\bibinfo {volume} {86}},\ \bibinfo {pages} {500} (\bibinfo {year}
  {2001})}\BibitemShut {NoStop}%
\bibitem [{\citenamefont {Ichikawa}\ \emph {et~al.}(2000)\citenamefont
  {Ichikawa}, \citenamefont {Uchida}, \citenamefont {Tranquada}, \citenamefont
  {Niem\"oller}, \citenamefont {Gehring}, \citenamefont {Lee},\ and\
  \citenamefont {Schneider}}]{ichi00}%
  \BibitemOpen
  \bibfield  {author} {\bibinfo {author} {\bibfnamefont {N.}~\bibnamefont
  {Ichikawa}}, \bibinfo {author} {\bibfnamefont {S.}~\bibnamefont {Uchida}},
  \bibinfo {author} {\bibfnamefont {J.~M.}\ \bibnamefont {Tranquada}}, \bibinfo
  {author} {\bibfnamefont {T.}~\bibnamefont {Niem\"oller}}, \bibinfo {author}
  {\bibfnamefont {P.~M.}\ \bibnamefont {Gehring}}, \bibinfo {author}
  {\bibfnamefont {S.-H.}\ \bibnamefont {Lee}}, \ and\ \bibinfo {author}
  {\bibfnamefont {J.~R.}\ \bibnamefont {Schneider}},\ }\href@noop {} {\bibfield
   {journal} {\bibinfo  {journal} {Phys. Rev. Lett.}\ }\textbf {\bibinfo
  {volume} {85}},\ \bibinfo {pages} {1738} (\bibinfo {year}
  {2000})}\BibitemShut {NoStop}%
\bibitem [{\citenamefont {Li}\ \emph {et~al.}(2011)\citenamefont {Li},
  \citenamefont {Alidoust}, \citenamefont {Tranquada}, \citenamefont {Gu},\
  and\ \citenamefont {Ong}}]{li11}%
  \BibitemOpen
  \bibfield  {author} {\bibinfo {author} {\bibfnamefont {L.}~\bibnamefont
  {Li}}, \bibinfo {author} {\bibfnamefont {N.}~\bibnamefont {Alidoust}},
  \bibinfo {author} {\bibfnamefont {J.~M.}\ \bibnamefont {Tranquada}}, \bibinfo
  {author} {\bibfnamefont {G.~D.}\ \bibnamefont {Gu}}, \ and\ \bibinfo {author}
  {\bibfnamefont {N.~P.}\ \bibnamefont {Ong}},\ }\href {\doibase
  10.1103/PhysRevLett.107.277001} {\bibfield  {journal} {\bibinfo  {journal}
  {Phys. Rev. Lett.}\ }\textbf {\bibinfo {volume} {107}},\ \bibinfo {pages}
  {277001} (\bibinfo {year} {2011})}\BibitemShut {NoStop}%
\bibitem [{\citenamefont {Karapetyan}\ \emph {et~al.}()\citenamefont
  {Karapetyan}, \citenamefont {{H\"ucker}}, \citenamefont {Gu}, \citenamefont
  {Tranquada}, \citenamefont {Fejer}, \citenamefont {Xia},\ and\ \citenamefont
  {Kapitulnik}}]{kara12}%
  \BibitemOpen
  \bibfield  {author} {\bibinfo {author} {\bibfnamefont {H.}~\bibnamefont
  {Karapetyan}}, \bibinfo {author} {\bibfnamefont {M.}~\bibnamefont
  {{H\"ucker}}}, \bibinfo {author} {\bibfnamefont {G.~D.}\ \bibnamefont {Gu}},
  \bibinfo {author} {\bibfnamefont {J.~M.}\ \bibnamefont {Tranquada}}, \bibinfo
  {author} {\bibfnamefont {M.~M.}\ \bibnamefont {Fejer}}, \bibinfo {author}
  {\bibfnamefont {J.}~\bibnamefont {Xia}}, \ and\ \bibinfo {author}
  {\bibfnamefont {A.}~\bibnamefont {Kapitulnik}},\ }\href@noop {} {\enquote
  {\bibinfo {title} {{Magneto-optical signatures of a cascade of transitions in
  La$_{1.875}$Ba$_{0.125}$CuO$_4$}},}\ }\Eprint
  {http://arxiv.org/abs/arXiv:1203.2977} {arXiv:1203.2977} \BibitemShut
  {NoStop}%
\end{thebibliography}

%

\end{document}